\documentclass[12pt,preprint]{aastex}
\usepackage{xspace} 
\usepackage{graphicx} 
\usepackage{natbib}
\usepackage{url}
\newcommand{\noopsort}[1]{}
\newcommand{\nraoblurb}{The National Radio Astronomy Observatory is
a facility of the National Science Foundation operated under cooperative
agreement by Associated Universities, Inc.}
\newcommand{\hide}[1]{}
\newcommand{\nexpo}[2]{\ensuremath{{#1}\times10^{#2}}\xspace}
\newcommand{\expo}[1]{\ensuremath{10^{#1}}\xspace}

\newcommand{\gsim}{\ensuremath{\,\gtrsim\,}\xspace}
\newcommand{\lsim}{\ensuremath{\,\lesssim\,}\xspace}
\let\phs\undefined
\newcommand\phs{\phantom{$-$}}%
\newcommand{\gl}{\ensuremath{\ell}\xspace}
\newcommand{\gb}{\ensuremath{{\it b}}\xspace}
\newcommand{\absb}{\ensuremath{\vert\,\gb\,\vert}\xspace}

\newcommand{\lb}{\ensuremath{(\gl,\gb)}\xspace}

\newcommand{\kms}{\ensuremath{\,{\rm km\,s^{-1}}}\xspace}

\newcommand{\microns}{\ensuremath{\,\mu{\rm m}}\xspace}

\newcommand{\cm}{\ensuremath{\,{\rm cm}}\xspace}

\newcommand{\kpc}{\ensuremath{\,{\rm kpc}}\xspace}
\newcommand{\K}{\ensuremath{\,{\rm K}}\xspace}
\newcommand{\mK}{\ensuremath{\,{\rm mK}}\xspace}

\newcommand{\ghz}{\ensuremath{\,{\rm GHz}}\xspace}

\newcommand{\degree}{\ensuremath{\,^\circ}\xspace}

\newcommand{\jy}{\ensuremath{\,{\rm Jy}}\xspace}

\newcommand{\mjy}{\ensuremath{\,{\rm mJy}}\xspace}

\newcommand{\te}{\ensuremath{{T_{e}}}\xspace}

\newcommand{\ev}{\ensuremath{\,{\rm eV}}\xspace}
\newcommand{\rgal}{\ensuremath{\,R_{\rm Gal}}\xspace}   

\newcommand{\hii}{{\rm H\,{\scriptsize II}}\xspace}

\newcommand{\he}[1]{\ensuremath{^{#1}{\rm He}}\xspace}

\newcommand{\her}[1]{\ensuremath{^{#1}{\rm He}/{\rm H}\xspace}}
\newcommand{\hepr}[1]{\ensuremath{^{#1}{\rm He}^{+}/\,{\rm H}^{+}\xspace}}
\newcommand{\hepp}[1]{\ensuremath{^{#1}{\rm He}^{++}\xspace}}
\newcommand{\heppr}[1]{\ensuremath{^{#1}{\rm He}^{++}/\,{\rm H}^{+}\xspace}}

\newcommand{\oiii}{{\rm O\,{\small III}}\xspace}

\newcommand{\hna}{\ensuremath{{\rm H}\,{\rm n}\,\alpha\xspace}}
\newcommand{\hnaa}{\ensuremath{\langle\,\hna\,\rangle\xspace}}

\shorttitle{H~{\small \bf II} Region Discovery Survey}
\shortauthors{Wenger et al.}

\begin{document}

\title{THE GREEN BANK TELESCOPE H{\small II} REGION DISCOVERY SURVEY:
IV. HELIUM AND CARBON RECOMBINATION LINES}

\author{Trey V. Wenger\altaffilmark{1}, T. M. Bania\altaffilmark{1},
Dana S. Balser\altaffilmark{2}, and L. D. Anderson\altaffilmark{3}}

\altaffiltext{1}{Astronomy Department, 725 Commonwealth Ave., Boston
  University, Boston MA 02215, USA.}  
\altaffiltext{2}{National Radio
  Astronomy Observatory, 520 Edgemont Road, Charlottesville VA,
  22903-2475, USA.}  
\altaffiltext{3}{Department of Physics, West
  Virginia University, Morgantown, WV 26506, USA.}

\begin{abstract}
The Green Bank Telescope \hii Region Discovery Survey (GBT HRDS) found
hundreds of previously unknown Galactic regions of massive star
formation by detecting hydrogen radio recombination line (RRL)
emission from candidate \hii region targets. Since the HRDS nebulae
lie at large distances from the Sun, they are located in previously
unprobed zones of the Galactic disk.  Here we derive the properties of
helium and carbon RRL emission from HRDS nebulae.  Our target sample
is the subset of the HRDS that has visible helium or carbon RRLs.
This criterion gives a total of 84 velocity components (14\% of the
HRDS) with helium emission and 52 (9\%) with carbon emission.  For our
highest quality sources, the average $\rm \he4^+/H^+$ abundance
ratio by number, $<y^+>$, is $0.068 \pm 0.023 (1\,\sigma)$.  This is the same
ratio as that measured for the sample of previously known Galactic
\hii regions.  Nebulae without detected helium emission give robust
$y^+$ upper limits.  There are 5 RRL emission components with $y^+$
less than 0.04 and another 12 with upper limits below this value.
These \hii regions must have either a very low \he4 abundance or
contain a significant amount of neutral helium.  The HRDS has 20
nebulae with carbon RRL emission but no helium emission at its
sensitivity level.  There is no correlation between the carbon RRL
parameters and the 8\,\microns mid-infrared morphology of these
nebulae.
\end{abstract}

\newpage

\section{Introduction\label{sec:intro}}
\hii regions are among the brightest radio sources in the Milky Way at
centimeter wavelengths. Within a molecular cloud, a massive newborn
early-type star (O4--B9) or cluster of such stars emits
extreme-ultraviolet (EUV: $h\nu >13.6$ eV) photons that ionize
hydrogen and helium in a surrounding zone around the stars. Beyond
this fully ionized plasma, the \hii region, is the photo-dissociation
region (PDR, sometimes also referred to as a ``photon dominated
region'') where less energetic photons can dissociate most molecules
and ionize atoms with ionization potentials below 13.6 eV, such as
carbon (11.3 eV), silicon (8.2 eV), and sulfur (10.4 eV).  The PDR
acts as a boundary between the fully ionized \hii region and the
neutral molecular cloud surrounding it. Because the stars that create
\hii regions have lifetimes of $\lsim$10 million years, they are zero
age objects compared to the age of the Milky Way. \hii regions thus
reveal sites of current or recent massive star formation.

Galactic \hii regions and their surrounding PDRs provide useful tools
for studying the Milky Way. They trace the current structure of the
Galaxy, reveal its chemical evolution history, and show the effect of
star formation on the evolution of the interstellar medium (ISM). \hii
regions have been well studied at optical wavelengths, primarily
because of the many bright collisionally excited lines \citep[e.g.,
  \oiii;][]{burke89,crawford00}, and optical recombination lines
(ORLs) \citep[e.g.,][]{peimbert74} that are available.  Optical
studies are limited by dust extinction, so they only probe a small
fraction of the Galactic disk.

Radio recombination lines (RRLs), however, are nearly extinction free
tracers of Galactic \hii regions. Because the Milky Way ISM is
optically thin at centimeter wavelengths, RRL emission can be detected
from nebulae located throughout the Galactic disk. RRL spectra of \hii
regions give the velocity which can be used to derive kinematic
distances to the nebulae \citep[e.g.,][]{anderson12}. With known
distances many nebular physical properties can be determined.
Distances are needed, for example, to study the \hii region luminosity
function and to trace Galactic structure \citep{downes80,anderson09}.
Furthermore, when combined with measurements of the thermal free-free
continuum emission from the \hii region, the nebular electron
temperature, $T_e$, can be derived from the line-to-continuum ratio.
The \hii region \te is a proxy for the nebular metallicity and can be
used to constrain chemical evolution models for the Milky Way
\citep{wink83,shaver83,quireza06b,balser11}.

Some \hii regions show RRL emission from heavier elements with
ionization potentials lower than that of hydrogen, such as carbon.
This emission comes from the PDR surrounding the \hii region where the
gas is denser and cooler \citep{hollenbach99}. Carbon RRL measurements
can help constrain the physical properties of the PDR.  They provide
information about the PDR kinematics which, when combined with \hii
region and molecular cloud velocities, provides insight into the
dynamics of regions forming massive stars \citep{roshi05}.  By
combining observations of carbon RRL and
158\,\micron\ C\thinspace{\small II} emission, the PDR density and
temperature can be derived \citep{natta94}.  Finally, \citet{roshi07}
suggested that the non-thermal carbon RRL line width is caused by
magnetic turbulence and is a measure of the magnetic field strength of
the PDR.

The Green Bank Telescope \hii Region Discovery Survey (hereafter GBT
HRDS) discovered hundreds of previously unknown Galactic \hii regions
by detecting RRL emission from these nebulae with the GBT at X-band (9
\ghz, 3 \cm).  Some HRDS spectra had emission components at several
different LSR velocities.  Altogether, the HRDS detected 603 hydrogen
RRLs from 448 \hii regions.  These nebulae are located in the
\lb\ zone $67\degree~\ge~\gl~\ge~343\degree,~\absb~\le~1\degree$.  The
HRDS census doubled the number of known \hii regions in this part of
the Galaxy.  \citet[][hereafter Paper~I]{bania10} described the GBT
HRDS survey.  \citet[][hereafter Paper~II]{anderson11} gave the GBT
HRDS methodology and provided the HRDS source catalog.
\citet[][hereafter Paper~III]{anderson12} derived kinematic distances
to a large fraction of the GBT HRDS sample.

Here we derive the helium and carbon RRL properties for a subset of
the HRDS nebulae. We choose the sample targets based on a visual
inspection of the HRDS spectra (Section~\ref{sec:sample}).  We then
fit Gaussians to the helium and carbon emission to determine the best
fit line parameters in Section~\ref{sec:analysis}.  We discuss these
results in Section~\ref{sec:disc} where we compare the helium RRL
properties of HRDS nebulae with the general population of Galactic
\hii regions in Section~\ref{sec:comparison} and their carbon RRL
properties with the population of Galactic PDRs in
Section~\ref{sec:pdrs}.  We identify 20 nebulae where carbon RRL
emission is detected in the absence of helium emission and
in Section~\ref{sec:morphology} find that no carbon RRL property seems
to be related to the nebular PDR morphology seen in 8\,\microns
images.  Using the HRDS helium recombination line parameters, we
derive in Section~\ref{sec:disc_abundance} the $\rm \he4^+/H^+$
abundance by number, $y^+$, and find a slight increase in $y^+$ as a
function of Galactocentric radius.  Finally, in Section~\ref{sec:disc_limits}
we use HRDS nebulae without detected helium emission to set $y^+$
upper limits and find that some \hii regions have a very low $y^+$
abundance.  

%

\section{Sample Selection\label{sec:sample}}
Sensitive measurements of hydrogen, helium, and carbon RRL parameters
are needed in order to derive accurate helium ionic abundance ratios
and to characterize the properties of the carbon emission from the
PDRs.  Our targets need to be that subset of the HRDS nebulae with the
most reliable helium and carbon detections.  We make a visual
inspection of the entire HRDS catalog and identify all sources with
visible helium emission.  Altogether, we identify 84 HRDS hydrogen RRL
emission components that also have helium emission strong enough to
derive accurate line parameters.

To ensure that our target sample is complete, we search the HRDS
catalog automatically, seeking those nebulae that ought to show
detectable helium emission based on the strength of the hydrogen RRL.
For their sample of Galactic \hii regions, \citet[hereafter
  QEAa]{quireza06a} found the average ratio of He/H RRL intensities at
3\cm wavelength to be $\rm T_{\rm L}({\rm He})/T_{\rm L}({\rm H}) =
0.096$.  Using this ratio together with the observed hydrogen RRL
intensity, we can predict which HRDS nebulae should show helium RRL
emission at the 2$\sigma$ intensity level if they are ``typical'' QEAa
\hii regions.  Here, the 1$\sigma$ level is individually set by the
rms spectral noise of each HRDS source. This 2$\sigma$ intensity
criterion, combined with the constraint that the helium emission be
found at the same velocity as that of the hydrogen RRL, identifies all
targets in the HRDS that can possibly have helium RRL emission.
These helium detection criteria identify a total of 163 sources.  They
find the same targets as the visual inspection does, together with an
additional 79 nebulae that might show helium RRL emission at the
2\,$\sigma$ level.  We find in Section~\ref{sec:analysis} that none of
these new targets has helium RRL emission that can be measured
accurately.  We use this sample of helium non-detections in
Section~\ref{sec:disc_limits} to derive upper limits on $y^+$ for
these nebulae.

Our sample is thus comprised of 84 hydrogen RRL components that have
helium RRL emission.  This is $\sim$14\% of the RRL velocity
components listed in the HRDS catalog.  Finally, we re-examine the
HRDS catalog to identify any sources with visible carbon RRL emission
that might be missed because our selection process is focused on the
helium emission.  Carbon RRL emission is seen in 52 targets (9\% of
HRDS) and there are 20 sources (3\% of HRDS) with carbon emission but
no detected helium line.

\section{Radio Recombination Line Properties\label{sec:analysis}}
To get the best possible hydrogen, helium, and carbon RRL parameters,
we reanalyze the spectra from Paper II. The analysis in Paper II
focused solely on getting accurate hydrogen RRL velocities. Those
baseline models are not optimized for an accurate analysis of helium
and carbon RRL parameters.  Here we reanalyze all 163 targets we find
using the automatic search.  The first step is to fit a new baseline
model for the spectra. We then fit a Gaussian function to the hydrogen
line. If there are multiple or blended hydrogen lines, we iterate to
get the best possible fit.

\begin{figure}
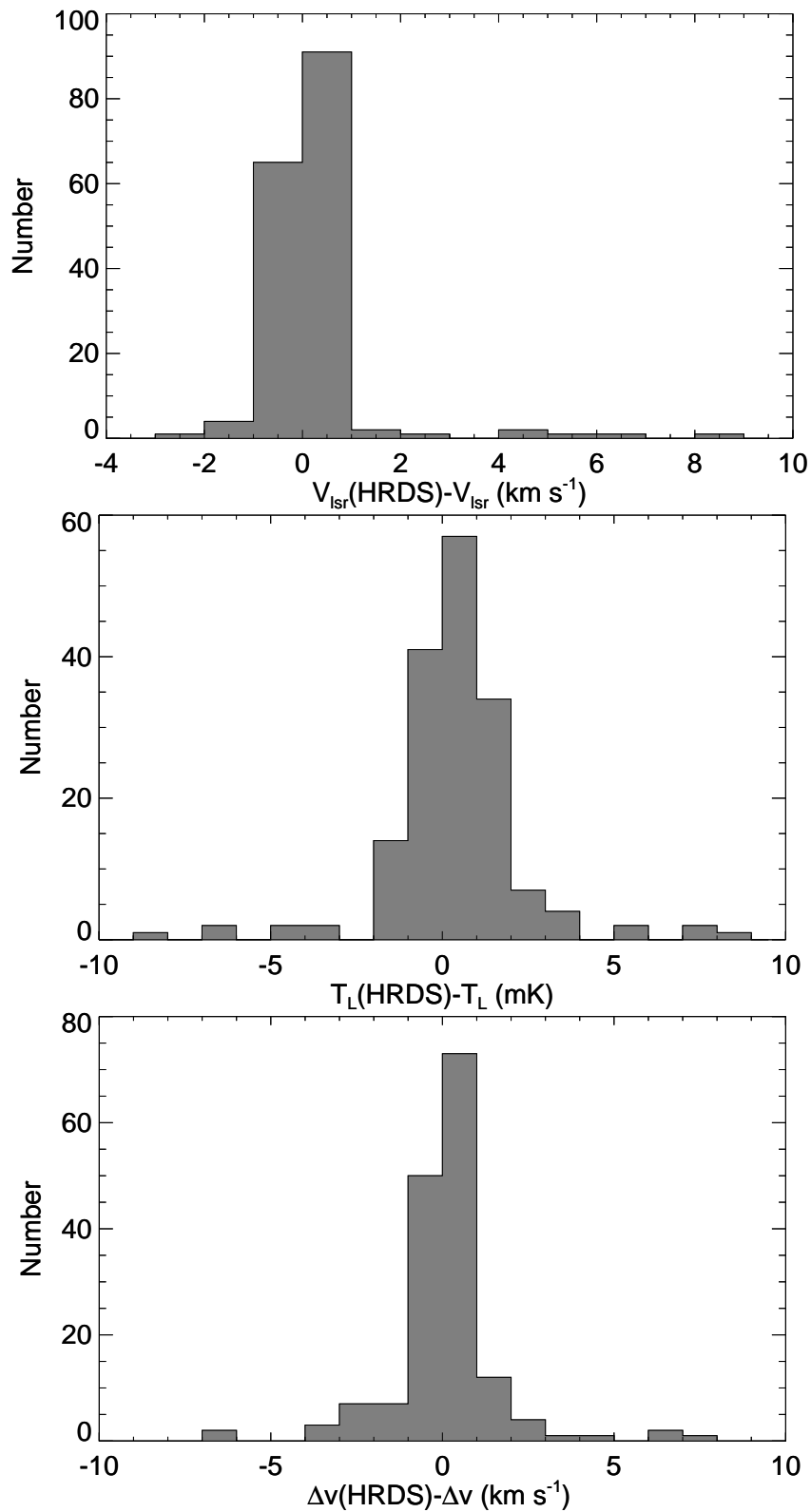

\centering
\includegraphics[scale=0.45,angle=90]{compare_center.epsi}
\includegraphics[scale=0.45,angle=90]{compare_peak.epsi}
\includegraphics[scale=0.45,angle=90]{compare_width.epsi}
\caption{Comparison of the HRDS catalog hydrogen RRL parameters with
  this reanalysis. Histograms show the differences between the
  Gaussian fitted line center (top), line intensity (middle), and FWHM
  line width (bottom). The outliers in each represent the same set of
  nebulae with poor fits in the original HRDS catalog.}
\label{fig:compare}
\end{figure}

Overall, the new baseline models and fits to the hydrogen lines
reproduce the original hydrogen RRL parameters in the HRDS
catalog. Figure~\ref{fig:compare} shows histograms comparing this
reanalysis with that of the HRDS for the sample of 163 targets.  It
makes this comparison for the hydrogen RRL LSR velocity (top), line 
intensity (middle), and FWHM line width (bottom).  From 
these data we find the mean difference in the hydrogen line LSR
velocity to be \lsim 0.1 \kms with a dispersion $\sim\,$0.4 \kms. The mean
difference in line intensities is $\sim$0.4 \mK with a dispersion
of $\sim\,$2  \mK.  The FWHM line width difference is also small with
a mean \lsim 0.1 \kms and a dispersion $\sim\,$1.0 \kms.
These mean differences for the latter two quantities are both
$\lsim 0.5\,\%$ of their average values for HRDS nebulae (Paper II).
Figure~\ref{fig:compare} is scaled for clarity.  We thus omit a few
outlier data points in these comparison plots: one each from the line
velocity and intensity comparisons and seven from the FWHM line
width comparison.  Each outlier is a source with multiple hydrogen
lines and had a poor fit in the original HRDS catalog.  

\begin{figure}
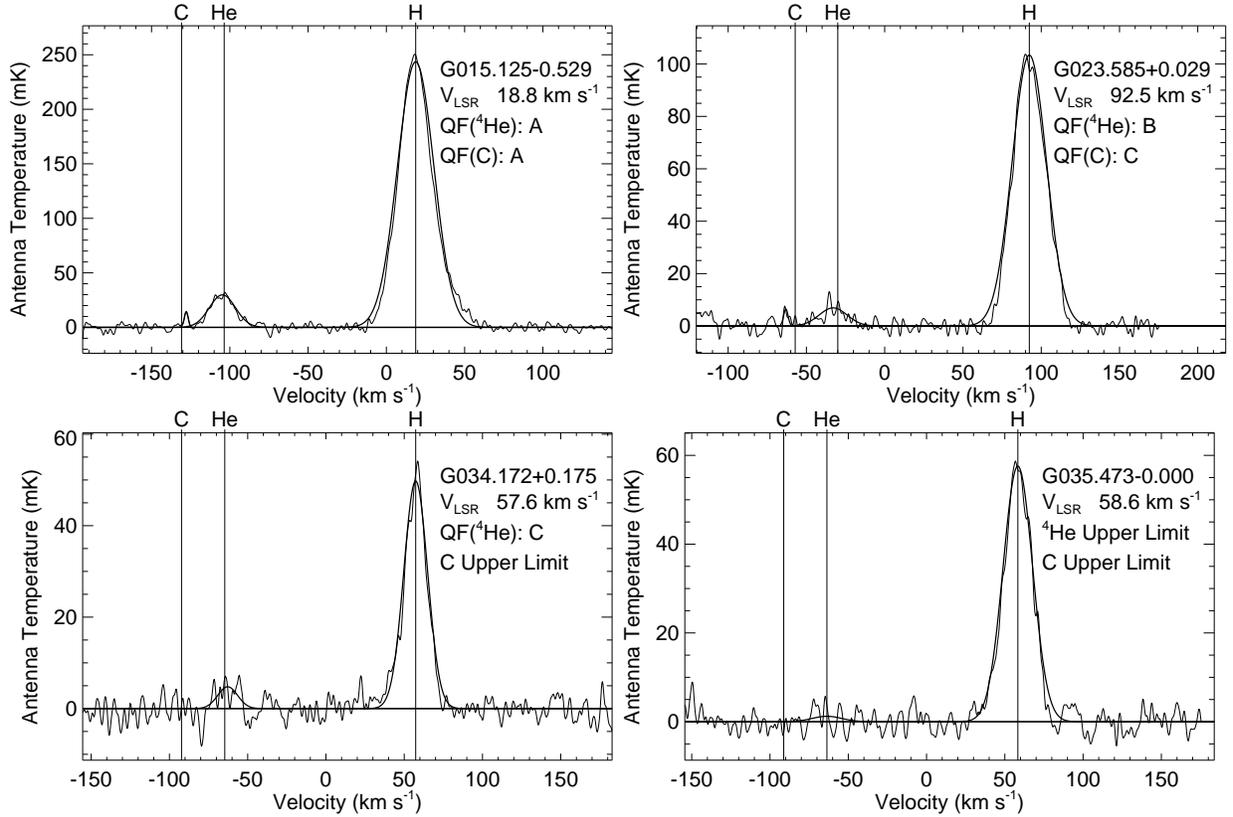

\centering
\includegraphics[scale=0.33,angle=90]{G015.125-0.529.epsi}
\includegraphics[scale=0.33,angle=90]{G023.585+0.029.epsi}
\includegraphics[scale=0.33,angle=90]{G034.172+0.175.epsi}
\includegraphics[scale=0.33,angle=90]{G035.473-0.000.epsi}
\caption{Example Gaussian fits to hydrogen, helium, and carbon
  recombination lines. Typical spectra are shown for different quality
  factors (QF; see text).  Vertical flags mark the RRL transitions.
  The positions of the helium and carbon line flags are based on
  the hydrogen velocity and are not fits.  The best fit Gaussians to
  the RRL emission are drawn as solid black lines.}
\label{fig:qf_examples}
\end{figure}

Next, we attempt to fit for helium emission.  We deem the helium fit
to be a detection if the fitted intensity, $T_L({\rm He})$, is greater
than twice the spectral rms noise, $2\sigma$, and the helium velocity
matches that of the hydrogen RRL component.  If there are multiple
hydrogen lines then we also fit for multiple helium lines.  We find 84
such helium RRL emission components in our sample of 163 candidates.
These are, in fact, the components we identify from a visual
inspection of the HRDS spectra.  Figure~\ref{fig:qf_examples} gives
examples of typical RRL spectra and the Gaussian fits.

We give the hydrogen and helium line parameters for these 84 RRL
components in Tables~\ref{tab:hdata} and \ref{tab:hedata},
respectively.  Source names with an appended letter,
e.g. ``G000.382+0.017a'', stem from the Paper II catalog and refer to
a specific emission component in a nebular spectrum that has hydrogen
RRLs at several different LSR velocities.  Each table lists the
intensity, $T_L$, the full width at half-maximum (FWHM) line width,
$\Delta v$, the LSR velocity, $V_{\rm LSR}$, the fit uncertainties,
e.g., $\sigma\, \Delta v$, and the rms spectral noise in the baseline
regions. The tables also contain a quality factor, QF, assigned by eye
to each fit. The QF is a qualitative judgment of the goodness of the
fit based on the accuracy of the baseline model, the rms spectral
noise, and the presence of multiple or blended lines which are often
difficult to fit. This is the same procedure as that used by QEAa and
QEAb (see these papers for more details). Of the 84 hydrogen velocity
components in our sample, we assign 82\% as QF A, 12\% as QF B, and
6\% as QF C.  For the helium fits, we rank 27\% as QF A, 32\% as QF B,
and 40\% as QF C. Once again, Figure~\ref{fig:qf_examples} gives
examples of typical spectra for each QF.  Our highest quality QF A and
B sources are the RRL components that should be used in any
quantitative analysis.

\begin{figure}
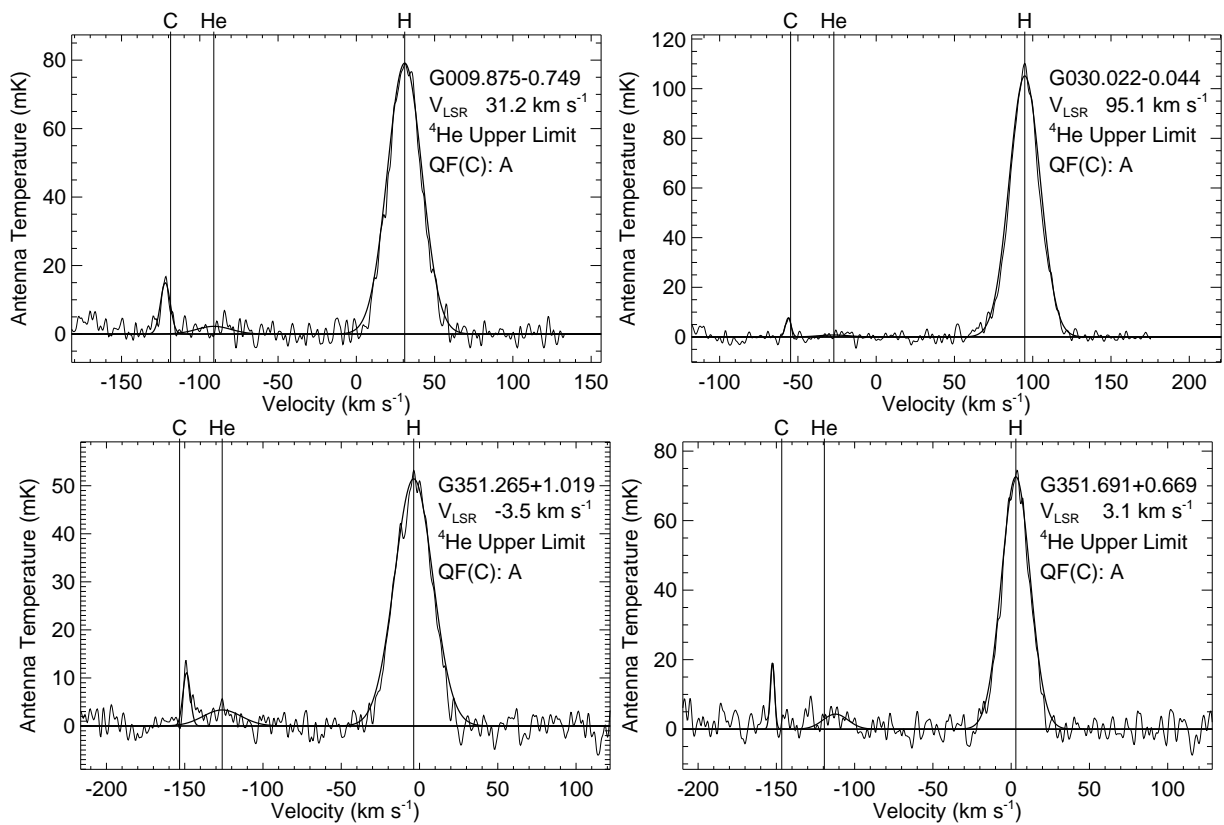

\centering
\includegraphics[scale=0.33,angle=90]{G009.875-0.749.epsi}
\includegraphics[scale=0.33,angle=90]{G030.022-0.044.epsi}
\includegraphics[scale=0.33,angle=90]{G351.265+1.019.epsi}
\includegraphics[scale=0.33,angle=90]{G351.691+0.669.epsi}
\caption{Example HRDS nebulae with carbon RRLs. No 2$\sigma$ helium
  detection is seen in these spectra; the Gaussian fits at the
  location of the helium flags are upper limits.}
\label{fig:carbon_lines}
\end{figure}

Since carbon RRL emission comes from the PDR and not the \hii region,
we do not expect the carbon velocity to match that of the hydrogen and
helium RRLs.  This can lead to blending of the carbon and helium RRL
emission.  Therefore, before we fit a Gaussian function to the carbon
emission, we subtract the helium emission from the spectrum using the
best fit Gaussian parameters from Table~\ref{tab:hedata}.  As was
shown by QEAa, this gives more accurate carbon line parameters.
Table~\ref{tab:cdata} lists the carbon fit parameters for the 52
emission components where a carbon line can be fit at the 2$\sigma$
level.  It contains the same information as Tables~\ref{tab:hdata} and
\ref{tab:hedata}.  Of the 52 carbon RRLs in our sample, we rank 42\%
as QF A, 40\% as QF B, and 17\% as QF C. Altogether, there are 20
nebulae that show carbon emission but no measurable helium emission
and 19 are QF A or B. Figure~\ref{fig:carbon_lines} shows a subset of
these sources which are discussed in Sections~\ref{sec:pdrs} and
\ref{sec:morphology}.  Table~\ref{tab:hdata} gives the hydrogen line
parameters for these nebulae.  We do not consider carbon upper limits
in this analysis.

\section{Discussion\label{sec:disc}}
\subsection{HRDS Nebulae and the Population of Galactic H\thinspace{\small II} Regions\label{sec:comparison}}
Compared to the sample of previously known \hii regions, the HRDS
nebulae are, on average, farther from the Sun and, at 3 \cm
wavelength, are both smaller in angular size and have weaker RRL and
continuum emission (Papers I, II, and III).  Here we examine whether
the RRL emission properties of HRDS nebulae are consistent with them
being part of the general population of Galactic \hii regions.  We
confirm, as one might expect, that the HRDS nebulae are typical
Galactic \hii regions.

Because the HRDS is the only RRL survey yet made with the GBT, some
care must be taken when comparing the HRDS with other recombination
line surveys.  These used other telescopes that typically have larger
beam sizes, lower gains (expressed as Kelvins of antenna temperature
per Jansky [K $\jy^{-1}$]), and, usually, poorer survey sensitivity
(i.e. spectral rms).  The most robust comparisons come from the
observed ratios of RRL parameters and the line-to-continuum
intensities because ratios minimize systematic effects caused by
different telescopes and spectrometer systems.  The most important
such effect is that, because of their different X-band beam sizes,
telescopes measure RRLs from varying volumes of ionized gas.  Here we
use \hepr4 ratios of RRL intensity and FWHM line width to compare the
HRDS with previously known \hii regions.  Analyses of the HRDS nebular
line-to-continuum ratios and the electron temperatures that can be
derived from these data will be the subject of a future paper in this
series (Balser, et al. 2013, in preparation).

Interpreting the ratios measured by different telescopes does require,
however, that one assume co-spatial ionized zones and uniform
properties throughout these volumes for all ionic species considered.
\citet{osterbrock89} showed that, for a \her4 abundance of 0.10, the
helium and hydrogen Str\"{o}mgren zones are co-spatial for all single
stars with effective temperatures greater than $\sim\,$39,000 \K .
These are stars of spectral type $\sim\,$O8 V or earlier
\citep{sternberg03}. The carbon RRL emission comes from the PDR and is
clearly not co-spatial with the helium and hydrogen Str\"{o}mgren
zone.  So, although carbon RRL parameter ratios may give some
empirical insight when comparing different RRL surveys, they have no
obvious astrophysical interpretation.  That the RRL properties of all
species are uniform throughout the nebular plasma is problematic as is
the requirement that the plasma fills the beams of all the telescopes
involved.  Nonetheless, ratios of observed RRL parameters are the best
available means of comparison.

We use the QEAa survey as a proxy for the properties of RRL emission
from the general Galactic \hii region population.  This X-band survey
was made with the NRAO 140 Foot telescope by members of the HRDS team.
It is comprised of RRL spectra toward 119 directions for a sample of
106 \hii regions.  Together, the QEAa nebulae are the brightest \hii
regions in the Northern sky.  With a typical rms spectral noise
ranging between 2.5 and 8.75 \mjy, the QEAa survey had at the time the
most sensitive RRL spectra ever made.  With an X-band gain of 2 \K
$\jy^{-1}$, however, the GBT is five times more sensitive than the 140
Foot (0.4 \K $\jy^{-1}$) for point sources.  The typical spectral rms
for HRDS nebulae is $\sim\,$1 \mjy (Paper II).  Furthermore, the 140
Foot's X-band FWHM beam for the QEAa RRL spectra is $\sim\,192$
\arcsec, whereas for the GBT's HRDS spectra it is $\sim\,82$ \arcsec.
The QEAa RRL parameters are therefore derived from an area on the sky
$\sim\,5.5$ times larger than that sampled by the HRDS.

\begin{figure}
\centering
\includegraphics[scale=0.70,angle=90]{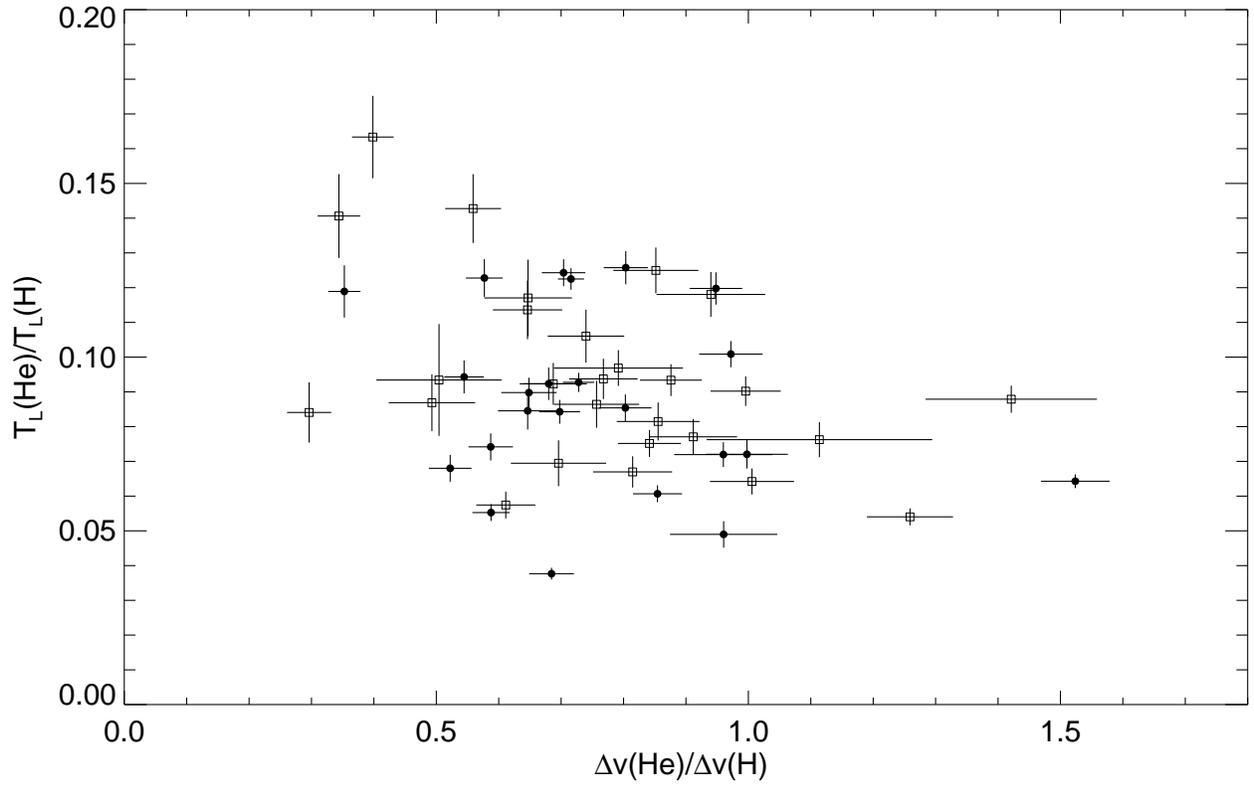}
\caption{Helium radio recombination line properties. Shown are the
  He/H ratios of the RRL intensities plotted as a function of FWHM
  line widths.  Only quality factor A (filled circles) and B (open
  squares) emission components are shown. Error bars are
  $\pm\,1\,\sigma$ uncertainties in the Gaussian fits.}
\label{fig:peak_width}
\end{figure}

The RRL parameters of the helium and carbon emission detected in the
HRDS are summarized in Tables~\ref{tab:hedata} and \ref{tab:cdata}.
Hereafter, we restrict our analyses to the highest quality emission
components, QFs A and B, a sample that includes 54 helium and 43
carbon emission lines.  We compare here these parameters to the best
RRL measurements of QEAa (QFs A, B, and C in that paper).
Figure~\ref{fig:peak_width} summarizes the helium RRL properties for
this restricted sample. It plots the ratios of the helium and hydrogen
RRL parameters: the relative line intensity, $\rm T_L(He)/T_L(H)$, as
a function of relative FWHM line width, $\rm \Delta v(He)/\Delta
v(H)$.
Based on previous RRL studies of Galactic \hii regions (QEAa is just
one example), we would not expect a correlation between the RRL
intensities and FWHM line widths. (The \citealt{lockman89} and
\citealt{lockman96} RRL surveys do show a weak correlation between
very low intensities and wide line widths, but these line parameters
are suspect due to spectral noise.  They would be QF C sources, or
upper limits, in the HRDS.)  As expected, Figure~\ref{fig:peak_width}
shows no correlation between the He/H ratios of RRL intensities and
line widths.

We summarize the average properties of the HRDS helium and carbon RRLs
in Table~\ref{tab:rrlproperties} which shows the mean and standard
deviation for the line intensity ratio, $\rm T_L(X)/T_L(H)$, the FWHM
line width ratio, $\rm \Delta v(X)/\Delta v(H)$, the LSR velocity
difference, $\rm V_{lsr}(X)-V_{lsr}(H)$, and the absolute value of the
velocity difference, $\rm \vert\,V_{lsr}(X)-V_{lsr}(H)\,\vert$.
Here, X is the subset of our sample being analyzed, i.e., the helium
or carbon RRLs.  We plot the distribution of these same quantities in
Figure~\ref{fig:He_params} for the helium RRLs and in
Figure~\ref{fig:C_params} for carbon.

The mean properties of the helium RRL emission from HRDS nebulae are
identical to those found by QEAa for their \hii region sample.
Comparing HRDS with QEAa shows that the line intensity and line width
ratios differ by less than $2\,\%$.  As did QEAa, we find that both
ratio distributions have substantial intrinsic width, and neither is
well described by a Gaussian.  The standard deviations of the ratio
distributions are also virtually identical.  Moreover, the magnitude
of these common standard deviations exceeds the difference between the
average ratios for the two nebular samples by a factor of ten.

We find that there is a real paucity of nebulae with $\rm
T_L(He)\,/\,T_L(H)\,<\,0.05$.  QEAa had high enough sensitivity to
detect such sources but also found very few.  The HRDS, with even
higher sensitivity, has only two more.  Nebulae such as these must be
relatively uncommon. 
As did QEAa, we find that the line width distribution drops off
sharply for values $\lsim$0.5 and $\gsim$1.  \hii region RRL line
widths arise from thermal broadening and turbulence.  In the absence
of any turbulence, the He/H line width ratio should be 0.5 due to the
mass effect alone, so the \hii regions where this ratio exceeds unity
either must be very highly turbulent or their helium and hydrogen
Str\"{o}mgren zones are not cospatial.

We compare the helium and hydrogen RRL velocities in two ways by
considering the LSR velocity difference, $\rm V_{lsr}(He)-V_{lsr}(H)$,
and the absolute value of the velocity difference, $\rm
\vert\,V_{lsr}(He)-V_{lsr}(H)\,\vert$.  The distribution of absolute
value differences is a better measure of systematic flows between the
hydrogen and helium plasmas.  We do not expect to see such flows,
however, because the HRDS \hii regions should have, on average,
cospatial Str\"{o}mgren zones that share the same kinematics.  As did
QEAa, we find no significant difference in the helium and hydrogen RRL
kinematics.  This is noteworthy given the disparate sizes of the
telescope beams.  The average of both the HRDS and QEAa \hii region
samples gives a velocity difference $\sim\,-0.2\,\kms$ and an absolute
value of the velocity difference is $\sim\,1.2\,\kms$.  For HRDS
nebulae, the mean velocity difference is $\sim\,$0.2\kms smaller than
that found by QEAa and the mean absolute value of the velocity
difference is $\sim\,$1.0\kms larger.  The absolute value of the
velocity difference distribution, however, is not Gaussian.  There are
four sources where this quantity is $\gsim\,5\,\kms$, suggesting
systematic flows between the helium and hydrogen gas.  Again, for
these few nebulae, the two Str\"{o}mgren zones may not be cospatial.

In sum, from these helium RRL comparisons of HRDS and QEAa nebulae, we
conclude that these objects all stem from a common population of
Galactic \hii regions.  This is further explored in
Section~\ref{sec:disc_abundance}.

\subsection{HRDS Nebulae and Galactic Photon Dominated Regions\label{sec:pdrs}}
Carbon RRLs provide important, complementary information about the
physical properties and kinematics of the PDRs surrounding \hii
regions.  Since the cosmic C/H abundance ratio by number is
$\sim\,\nexpo{3}{-4}$ \citep{lodders03,caffau10}, carbon RRL emission
from PDRs should be weak given the expected carbon column densities.
Naively, one might also expect to see carbon emission from PDRs in
front of, and behind, the \hii region.  \citet{roshi05}, however, used
multi-frequency data to constrain non-LTE models for PDRs.  They
concluded that the carbon RRL emission at X-band is dominated by
stimulated emission.  Their models showed that, because of stimulated
emission, the carbon RRL emission from the foreground PDR material is
five times brighter than the background PDR emission.  The HRDS carbon
RRL spectra thus preferentially measure emission from PDR material
that is both outside, and also in front of, the \hii region plasma
emitting thermal free-free continuum photons.  The properties of \hii
region carbon RRL emission from PDRs are therefore not expected to
match those of RRLs from the hydrogen and helium Str\"{o}mgren zones.

The average carbon RRL properties for HRDS \hii regions are summarized
in Table~\ref{tab:rrlproperties} and histograms of the distribution of
C/H ratios for these properties are shown in
Figure~\ref{fig:C_params}.  As was also the case for QEAa carbon RRL
emission, the histograms show that these distributions are not
well-described by Gaussians.  This is probably because not only is the
PDR material inhomogeneous, but also the PDR/\hii line of sight
geometry must be favorable for stimulated emission to occur.
Except for the mean C/H line intensity ratio, all other HRDS carbon
RRL parameters are consistent with those found by QEAa for their \hii
region sample.  The HRDS nebular means for the C/H line width ratio,
together with the carbon and hydrogen RRL LSR velocity difference, and
absolute velocity difference, are all within $\sim\,1\,\%$ of the QEAa
values.

The mean HRDS C/H line intensity ratio is $\sim\,$4 times greater, and
the dispersion of this quantity is $\sim\,$7 times greater, than what
QEAa found.  This result stems from the many HRDS nebulae that have
strong carbon RRL emission, often in the absence of any helium RRL
emission.  We find 19 QF A or B nebulae that have carbon, but no
helium, emission.  QEAa found only one such object and it had a poor
QF.  We show some of these \hii regions in
Figure~\ref{fig:carbon_lines}.  These nebulae are interesting because
their C/H line intensity ratios are large, with the mean $\rm
<T_L(C)/T_L(H)>\,=\,33.6\% \pm 33.9\%$ (the ``C with no He'' entry in
Table~\ref{tab:rrlproperties}).

If we eliminate these 19 nebulae and recompute the mean carbon RRL
parameters, we get the ``C with He'' values in
Table~\ref{tab:rrlproperties}.  Even this mean C/H line intensity
ratio is $\sim\,$2 times greater than what QEAa found.  We speculate
that the much smaller GBT beam must, on average, be more completely
filled with PDR material than is the 140 Foot beam.  We discuss the
six HRDS nebulae with the largest C/H intensity ratios in
Section~\ref{sec:morphology}.

Carbon RRL line widths also result from a combination of thermal and
turbulent broadening.  Because the PDR material is denser and colder
than the \hii region plasma, however, the carbon RRL line widths
should be narrower than the hydrogen and helium RRLs. The mass effect
alone would yield a carbon line width a factor of $\sqrt{12}\,\sim\,$
3.5 times narrower than the corresponding hydrogen line, giving
$\Delta\,V(C)\,/\,\Delta\, V(H)\,\sim\,0.29$.  The PDR gas is also
much colder, $\sim \expo{3}$ K in the $C^+$ zone, than the $\sim
\expo{4}$ K \hii region plasma.  This gives a further narrowing of the
carbon RRLs by a factor of $\sqrt{10}$, so $\Delta\,V(C)\,/\,\Delta\,
V(H)\,$ might be $\sim\,0.1$.  The average C/H line width ratio for
HRDS and QEAa nebulae together is $\sim\,0.32$ and both samples show a
significant number of nebulae with ratios greater than 0.4.  This
confirms that the carbon RRL line widths from PDRs are
non-thermal. Turbulence thus plays a major role in PDRs.
\citet{roshi07} suggested that the non-thermal carbon RRL line width
is caused by magnetic turbulence and is a measure of the magnetic
field strength of the PDR.

We do not expect the PDR and \hii region kinematics to be identical.
The carbon RRL and helium/hydrogen RRL velocity difference thus
directly probes the relative motion between the PDR and \hii region
gas.  We compare in Figure~\ref{fig:C_params} the carbon and hydrogen
RRL velocities in two ways by considering the LSR velocity difference,
$\rm V_{lsr}(C)-V_{lsr}(H)$, and the absolute value of the velocity
difference, $\rm \vert\,V_{lsr}(C)-V_{lsr}(H)\,\vert$.  Both the HRDS
and QEAa find that the velocity difference distributions are
$\sim\,10$ times wider than the same comparison for helium and
hydrogen RRLs from \hii regions.  Although the mean difference for
both \hii region samples is $\sim\,$ 0 \kms, the dispersions, $\sim\,$5
\kms, are much larger than this and they estimate the magnitude of
systematic flows between PDRs and \hii regions.  An even better
measure of such flows is the mean absolute value of the PDR/\hii
region velocity difference. The HRDS and QEAa samples find this to be
$\sim\,3-4\,\kms$.  Furthermore, using the Arecibo telescope at X-band
to measure RRL emission, \citet{roshi05} also found that the relative
motion between \hii region ionized gas and the associated PDR material
had an rms velocity difference of 3.3 km/s.  From all these data alone
we have no way of distinguishing between PDR gas flowing toward, and
gas flowing away from, the \hii regions. 

We are, however, clearly detecting carbon emission from PDR gas that
lies between us and the \hii region.  Moreover, the carbon line
intensity is enhanced due to stimulated emission (e.g.,
\citealt{roshi05}; QEAa).  For their \hii region sample, QEAa found a
strong correlation between the carbon RRL and the continuum
intensities (see their Fig.~11).  Here, in Figure~\ref{fig:tc_vs_th}
we plot instead the carbon RRL intensity as a function of the hydrogen
RRL intensity.  The hydrogen intensity is a proxy for the thermal
free-free continuum emission because the average line-to-continuum
ratio at 3 \cm wavelength for Galactic \hii regions is $\sim\,$0.1
(QEAa).  As did QEAa, we see a rough correlation between the carbon
RRL and continuum intensity which is what one expects if the carbon
intensity is being enhanced by stimulated emission.

For the HRDS nebulae, however, there are more \hii region outliers in
the top left hand corner (high carbon and low continuum intensity) of
Figure~\ref{fig:tc_vs_th} than are present in the QEAa \hii region
sample.  Most of these sources have Galactic longitudes within
15\degree of the Galactic Center.  Because of velocity crowding in
this region, we cannot determine accurate kinematic distances for
these nebulae (Paper III).  Nonetheless, they are located in the
direction of the innermost Galaxy which is known to have a substantial
non-thermal component to the total 3 \cm continuum emission.  We
speculate that, compared to PDRs that are excited by only thermal
emission from the \hii region plasma, the stimulated emission boost
for these outliers is further enhanced by an additional contribution
of non-thermal emission to the continuum.

In the HRDS we see more sources with bright carbon, but no helium, RRL
emission than in other \hii region samples (e.g, QEAa).
Interestingly, all of the outliers in Figure~\ref{fig:tc_vs_th} have
no helium RRL emission.  Is this significant?  We may not detect
helium emission because: (i) the nebular \her4 abundance is low; (ii)
the plasma is excited by later type stars and so helium is
under-ionized relative hydrogen; or (iii) the HRDS does not have
sufficient sensitivity to detect the helium emission.  Based on
previous He/H RRL line intensity ratios (e.g., QEAa, Eq.~2), we do not
expect to detect helium in about a third of these sources (the 7 open
rectangles in Fig.~\ref{fig:tc_vs_th}).  These are nebulae that lie
the farthest from the general trend that has the carbon RRL intensity
track the continuum intensity.  We see no helium RRL emission in these
nebulae because the HRDS lacks the sensitivity to do so.

\subsection{Carbon RRL Emission and H\thinspace{\footnotesize II} 
Region MIR Morphology  \label{sec:morphology}}
The best images of PDRs surrounding Galactic \hii regions come from
\emph{Spitzer} GLIMPSE 8.0\microns MIR data.  The 8.0\microns GLIMPSE
images trace PDRs because they include emission from polycyclic
aromatic hydrocarbon (PAH) molecules that fluoresce in ultraviolet
radiation fields.  A structure commonly seen in these PAH images is a
``bubble'' morphology; \citet{churchwell06,churchwell07} have
cataloged 420 of these objects.  In Paper II we found that 211 HRDS
nebulae are also \emph{Spitzer} GLIMPSE 8.0\microns bubble sources.
We suggested that this large number of HRDS bubble nebulae meant that
these objects are three-dimensional structures.  We also speculated
that all GLIMPSE bubbles are caused by \hii regions and that $\sim\,$
50\% of all Galactic \hii regions have a bubble morphology at
8.0\microns.  This motivated us to classify the 8.0\microns morphology
of all HRDS nebulae in an attempt to categorize the PDR/\hii region
geometry.  The qualitative, visual classifications we used were ``B''
for bubble, ``BB'' for bipolar bubble, ``PB'' for partial bubble,
``IB'' for irregular bubble, ``C'' for compact, ``PS'' for point
source, and ``ND'' for not detected (see Paper II for more details).

There are six HRDS nebulae with C/H line intensity ratios,
$T_L(C)/T_L(H)\,$, greater than 0.35.  The C/H line intensity ratio
histogram in Figure~\ref{fig:C_params} does not plot these six largest
ratios.  We do not see any obvious patterns of MIR morphological
peculiarities for these \hii regions, except for G056.252$-$0.160 and
G344.991$-$0.266.  Both have $T_L(C)/T_L(H)\,\gsim\,1$.  They are each
QF A nebulae and neither has any helium RRL emission.  Their MIR
morphologies are ``BB'' and ``C'', respectively.  In MIR three color
\emph{Spitzer} images, these nebulae seem to be interacting with
another PDR and in each case both sources are in the GBT beam (see
\url{go.nrao.edu/hrds}).  The strong carbon RRL emission may thus stem
from PDR material in the GBT beam that has higher density and lower
temperature than is typically the case.  Both \hii regions have very
low hydrogen RRL line widths, $\sim\,16 \kms$, indicating that these
nebulae have very cool plasmas and less turbulence than the general
Galactic \hii region population.  In fact, the average C/H line width
ratio for these nebulae is $\Delta\,V(C)\,/\,\Delta\,
V(H)\,\sim\,0.29$, which is precisely what one expects from the mass
effect alone.  

The strong carbon RRL emission that we see in some HRDS nebulae might
be due to a number of reasons.  The PDR material for HRDS nebulae is
more inhomogeneous than \hii region plasma.  Perhaps these strong
carbon lines stem from \hii regions where a larger volume of the PDR
than usual is measured by the GBT beam.  It is certainly true that for
stimulated emission to occur there must be sufficient carbon column
density and the PDR/\hii line of sight geometry must be favorable.
Whatever the case, perhaps the MIR PDR morphology of carbon emission
nebulae can provide some insight.

We test this hypothesis by analyzing the nebular carbon RRL properties
as a function of the MIR morphological classifications assigned in
Paper~II.  We compare the mean C/H ratios of various RRL properties
for different MIR morphological classifications in
Table~\ref{tab:c_morph}.  The table lists the nebular morphology, and
the number of carbon RRL emission components with that classification,
N, as well as the mean and standard deviation for the line intensity
ratio, $\rm T_L(C)/T_L(H)$, the FWHM line width ratio, $\rm \Delta
v(C)/\Delta v(H)$, the LSR velocity difference, $\rm
V_{lsr}(C)-V_{lsr}(H)$, and the absolute value of the velocity
difference, $\rm \vert\,V_{lsr}(C)-V_{lsr}(H)\,\vert$.  We only give
the morphological classification properties for QF A and B nebulae in
Table~\ref{tab:c_morph}.

Because the sample size for each morphological classification is
small, the standard deviations are quite large. This makes it
difficult to determine if there is a correlation between any of the
line parameters and the nebular MIR morphology.  One possibility is
that the nebulae with PDRs within the GBT beam should have the
strongest carbon emission.  The compact, ``C'' morphology, \hii
regions are only marginally resolved by the GLIMPSE survey, which has
a resolution of $2\,\arcsec$.  These nebulae are entirely within the
GBT beam.  We might then expect compact morphology sources to have, on
average, the strongest carbon lines.  In fact, the means and
dispersions we find for all the RRL parameters suggest that the RRL
properties and kinematics for nebulae of all morphological types are
indistinguishable.  This implies that individual source geometries may
be more important than 8.0\microns morphologies or that the Paper~II
classification criteria are not relevant for predicting the strengths
of carbon RRLs from PDRs.


\subsection{Helium Abundances \label{sec:disc_abundance}}

Measurements of the spatial distribution of elemental abundances
provide key constraints for our understanding of Galactic chemical
evolution.  The \hii region \he4 abundance relative to hydrogen by
number, $y\,=\,\her4$, is difficult to determine because one must know
the neutral and ionic abundance ratios, $y = y^0 + y^+ + y^{++}$,
where $y^0 = {\rm \he4^0/\,H^+}$, $y^+$ = \hepr4, and $y^{++}$ =
\heppr4.  (Recall that hydrogen in \hii regions is fully ionized so
the $\rm H^+$ abundance is the H abundance.)  Both RRLs and ORLs can
be used to study $y^+$ and $y^{++}$, but they cannot measure $y^0$
since there is no spectral transition for neutral helium.  Because the
high principal quantum number states of hydrogen and helium should
respond to radiative and collisional effects in the same way, their
line intensity ratio is equal to the abundance ratio. The
interpretation of RRLs is therefore much simpler than for ORLs.  RRLs
can be used to measure $y^+$ and $y^{++}$ directly, providing
information about the \hii region excitation
\citep[e.g.,][]{balser06}.

We derive the $\rm \he4^+/\,H^+$ abundance ratio by number from
the helium and hydrogen RRL properties:  
\begin{equation}
y^+=\frac{T_{\rm L}({\rm He})\ \Delta v({\rm He})}{T_{\rm L}({\rm
    H})\ \Delta v({\rm H})}
\label{eq:y}
\end{equation}
where $T_{\rm L}$ is the line intensity and $\Delta v$ is the FWHM
line width. Here we are using the area of the RRL emission set by a
Gaussian fit as a measure of the equivalent width.  Thus $y^+$
measures the $\rm \he4^+/\,H^+$ abundance ratio directly if the
source is optically thin.  The distribution of $y^+$ values that we
derive for HRDS nebulae is shown in Figure~\ref{fig:yp_hist}.  We find
that the mean abundance ratio by number is $<y^+>\, =\,0.068 \pm
0.023$, where the uncertainty is the 1$\sigma$ standard deviation of
the Figure~\ref{fig:yp_hist} distribution. This result is the same as
the abundance ratio found by QEAa, $<y^+>\,=\, 0.075 \pm 0.024$.

We find five sources with $y^+\,>\,0.10$. Enhanced $y^+$ values have
been measured in other \hii regions, such as W3A
\citep{roelfsema92,adler96}, W49A/M \citep{depree97}, G0.15$-$0.05
\citep{lang97}, NGC~6888 \citep{esteban92}, and NGC~6334A and K3-50A
\citep{balser01}.  There are some possible explanations for these high
$y^+$ measurements \citep[e.g.,][]{balser01}. We may be seeing an
enhancement of \he4 from helium rich material ejected by massive
stars.  Too, because of a hard radiation field (or other radiative
transfer effects), the helium Str\"omgren zone may be larger than that
of the hydrogen.  We may also have some observational or RRL fitting
errors due to blended carbon and helium RRLs. Blended helium and
carbon emission would increase the fitted width of the helium line and
thus increase the measurement of $y^+$.

Because HRDS \hii regions are located, on average, at the far
kinematic distance or beyond the Solar orbit (Paper III), they probe a
new zone of the Galactic disk compared to the sample of previously
known \hii regions.  We see a slight increase in $y^+$ with
Galactocentric radius, \rgal, for the HRDS nebulae.
Figure~\ref{fig:rgal} shows $y^+$ plotted as a function of \rgal for
QF A and B sources.  Here, there are no \hii regions inside \rgal
$\sim\,$4 \kpc because we were unable to derive accurate \rgal values
for 18 nebulae in the direction of the Galactic Center.  An unweighted
linear least-squares fit, which gives a slope of $0.0035\,\pm\,0.0016
\kpc^{-1}$, is shown in Figure~\ref{fig:rgal}.

Previous studies found that $y^+$ decreases or remains constant with
Galactocentric radius.  Since our study has only 5 nebulae beyond an
\rgal of 8 \kpc, a better analysis of Figure~\ref{fig:rgal} requires
more sophisticated numerical tools.  We use the program
SLOPES\footnote{See
  \url{http://www.astro.psu.edu/users/edf/research/stat.html}} to make
an ordinary least-squares regression using asymptotic error formulae
\citep{isobe90,feigelson92}.  These uncertainties for this type of fit
are valid for a large, $N\,\gsim\,$50, data sample but underestimate
the true error for a smaller sample.  Because of this, we use the
jackknife resampling procedure in SLOPES to derive more accurate
uncertainties.  This analysis gives exactly the same slope as
before, $0.00350\,\pm\,0.00160 \kpc^{-1}$, which says that there is a
positive gradient in $y^+$ where its value increases with \rgal.  This
is a weak result, however, since the Pearson correlation coefficient
for the Figure~\ref{fig:rgal} data is only 0.35 and the fit errors are
relatively large.

In any case, it is difficult to interpret the distribution of $y^+$
across the Galactic disk because we do not know either $y^{++}$ or
$y^0$ for most \hii regions.  Ordinary O-type Population I stars do
not radiate any appreciable number of photons with $h \nu\,>\,54.4
\ev$.  Galactic \hii regions are therefore not expected to have any
\hepp4 zone so $y^{++}$ should be zero.  Then, if there also were no
neutral helium, $y^0$ would also be zero and the \hepr4 value, $y^+$,
would measure \her4, the total helium abundance ratio by number, $y$.
But we do not know this to be the case.
Again, since there is no spectral transition for neutral helium, we
cannot directly determine $y^0$.  For some nebulae the ionization
structure can be inferred using optical transitions but there are only
a few \hii regions (e.g., M17, S206) where one expects there to be
little or no neutral helium \citep{carigi08,balser11}.  Care must
therefore be taken when interpreting, for example, the $y^+$ gradient
in the Galaxy.  Since \he4 is produced in stars, we expect a negative
gradient in the $y$ abundance as one moves radially outward in the
galactic disk.  The \her4 abundances should be higher for smaller
values of \rgal.  For $y^+$, however, the opposite may be the case.
Since the ISM metallicity is also expected to have a negative radial
gradient across the Galaxy, stars in the outer Galaxy should have
fewer metals and metals preferentially absorb higher energy photons.
Low metallicity stars thus produce harder radiation fields than high
metallicity stars.  \hii regions in the outer Galaxy might therefore
have higher values of $y^+$, resulting in a positive gradient. 

\subsection{Helium Abundance Upper Limits
\label{sec:disc_limits}}
Upper limits for the \hepr4 ionic helium abundances of \hii regions
can in principal be used to constrain the ionization state of these
nebulae.  Galactic \hii regions with a low upper limit on the
abundance, $y_{\rm limit}^+$, must have either a real lack of helium
or an insufficient number of 13.6 \ev photons to ionize their helium.
The latter nebulae must have a high neutral helium abundance.  We
define the $y_{\rm limit}^+$ limit for an \hii region with no visible
helium emission to be:
\begin{equation}
y_{\rm limit}^+ = 0.813\, \frac{2\,\sigma}{T_{\rm L}({\rm H})}.
\label{eq:ylimit}
\end{equation}
This limit stems from Eq.~\ref{eq:y} where here it is set by assuming
that the helium RRL line intensity is at most 2\,$\sigma$, where
$\sigma$ is the rms spectral noise of the spectrum, and $T_L({\rm H})$
is the intensity of the hydrogen RRL.  For the limit, we estimate the
He/H line width ratio to be the 0.813 average FWHM ratio, $\Delta
v({\rm He})/\Delta v({\rm H})$, measured by QEAa for their \hii 
region sample. 

The distribution of the \hepr4 upper limit abundances we derive for
HRDS nebulae with no visible helium emission is shown in
Figure~\ref{fig:all_limits}.  The black hatched histogram is the
$y_{\rm limit}^+$ distribution for the 79 nebulae with the most
astrophysically meaningful limits.  Unless their $y^+$ is anomalously
low, these \hii regions ought to have helium RRL emission at the
2$\,\sigma$ intensity level if they are ``typical'' Galactic \hii
regions (see Section~\ref{sec:sample}).  For this sample we find an
average upper limit of $<y_{\rm limit}^+>\,=\,0.15 \pm 0.14$ which is
two times larger than the average $y^+$ we find for the HRDS nebulae
that show helium emission.  The gray histogram shows the $y_{\rm
  limit}^+$ distribution for the remaining emission components.  (Here
we exclude the 7 emission components that have unphysical limits,
$y_{\rm limit}^+ > $1.)  For this sample we find an average upper
limit of $<y_{\rm limit}^+>\,=\,0.25 \pm0.23$.  A plot of the nebular
$y_{\rm limit}^+$ as a function of spectral rms noise shows no trend,
except for the expected increase in the size of the errors as the
spectral noise increases.


We show $y_{\rm limit}^+$ as a function of Galactocentric radius in
Figure~\ref{fig:ylimit_rgal}.  We plot only those nebulae where we
could determine an accurate \rgal from the Brand rotation curve.  This
eliminates 141 sources that lie in the direction of the galactic
center.  Black squares show the nebulae with the most astrophysically
meaningful limits and the open circles are the remaining limits for
HRDS nebulae.  The solid line marks the fit to our $y^*$
determinations (Sec.~\ref{sec:disc_abundance} and
Fig.~~\ref{fig:rgal}).  This fit shows that the actual $y^+$
measurements define a lower envelope for the $y_{\rm limit}^+$ upper
limits.
There are 5 HRDS \hii regions that have $y^+$ abundance ratios below
0.04 and another 12 with $y_{\rm limit}^+$ below 0.04.  These nebulae
must have either a very low \he4 abundance or a significant amount of
neutral helium.  The spectral noise of these low upper limit nebulae
spans the full range of HRDS sensitivity so they should be robust.
All of these \hii regions are located in the inner Galaxy at \rgal
\lsim 6 \kpc.

\section{Summary\label{sec:summary}}

\hii regions and PDRs provide information about the structure,
dynamics, and evolution of the Galaxy. We measure the line parameters
for 82 helium and 54 carbon RRL emission components in the HRDS
catalog.  For our highest quality \hii regions, we derive the average
properties of ratios of the helium and carbon emission line parameters
relative to that of hydrogen.  The HRDS \hii region sample includes 20
nebulae that show carbon, but no helium, RRL emission. Some of these
nebulae show extremely strong carbon emission.  Since the carbon RRLs
stem from stimulated emission in the foreground PDR material, nebulae
with carbon emission should have line of sight geometries where the
foreground PDR material covers a significant fraction of the continuum
emission from the \hii region plasma.  There does not seem to be any
correlation, however, with the carbon RRL line parameters and the \hii
region PDR morphology as seen in \emph{Spitzer} GLIMPSE 8\,\microns
images.

We derive the \hepr4 ionic abundance ratio by number, $y^+$, for HRDS
\hii regions using the measured line parameters.  The mean abundance
ratio, $<y^+>\,=\, 0.068 \pm 0.007$, is the same as that found by
\citet{quireza06a} for a sample of typical Galactic \hii regions.  In
contrast to all previous studies, however, the distribution of HRDS
\hii region \hepr4 abundance ratios shows a slight increase in
$y^+$ with Galactocentric radius.  It is difficult to interpret this
result because we do not know either $y^{++}$ or $y^0$ for these
nebulae.  Finally, we derive upper limits for the \hepr4 abundance
from nebulae without helium emission.  There are 5 RRL emission
components with $y^+$ less than 0.04 and another 12 with upper limits
that are also below this value.  These \hii regions have either a very
low \he4 abundance or contain a significant amount of neutral helium.
All these nebulae are located in the inner Galaxy at Galactocentric
radii less than 6 \kpc.

Because HRDS \hii regions are located, on average, at the far
kinematic distance or beyond the Solar orbit, they probe a new zone of
the Galactic disk compared with the sample of previously known \hii
regions.  We find, however, that the average properties of helium and
carbon RRL emission from HRDS nebulae are consistent with those found
for the general population of Galactic \hii regions by previous
studies.  That the average RRL properties match, however, does not
mean that there are not patterns to be found in their distribution
across the Galactic disk. In a future paper in this series, we shall
combine measurements of the thermal free-free continuum and
recombination line emission to derive the nebular electron
temperature, \te, from the line-to-continuum ratio. Because the \hii
region \te is a proxy for the nebular metallicity it can be used to
constrain chemical evolution models for the Milky Way (e.g.,
\citealt{balser11}).  The HRDS nebulae will allow us to search for
radial and azimuthal metallicity structure in a new zone of the
Galactic disk.

\appendix
\section{The HRDS Web Site}
All the helium and carbon RRL data are now incorporated in our HRDS
Web site, \url{http://go.nrao.edu/hrds}.  This site allows one to view
{\em Spitzer} three-color images and the \hnaa\ recombination line
spectra.  One can also download all the RRL data. As we continue to
extend the HRDS all future data will also be available on this site.
 
\begin{acknowledgments}
\nraoblurb
The HRDS was partially supported by NSF award AST 0707853 to TMB.  LDA
was partially supported by the NSF through NRAO GSSP awards 08-0030
and 09-005 from the NRAO.  TVW was partially supported by the NSF
through NRAO Summer REU award AST 1062154.
 
\em {Facility: Green Bank Telescope}

\end{acknowledgments}

\bibliography{ref.bib}
\bibliographystyle{apj}

\clearpage
\centering
\begin{deluxetable}{lrrrrrrrc}
\tablecaption{Hydrogen Radio Recombination Line Parameters}
\tablewidth{0pt}
\tablehead{
  \colhead{Source} &
  \colhead{$T_{\rm L}$} &
  \colhead{$\sigma T_{\rm L}$} &
  \colhead{$\Delta v$} &
  \colhead{$\sigma \Delta v$} &
  \colhead{$V_{\rm lsr}$} &
  \colhead{$\sigma V_{\rm lsr}$} &
  \colhead{rms} &
  \colhead{QF} \\

  \colhead{ } &
  \colhead{\mK} &
  \colhead{\mK} &
  \colhead{\kms} &
  \colhead{\kms} &
  \colhead{\kms} &
  \colhead{\kms} &
  \colhead{\mK} &
  \colhead{ }
}

\startdata
G000.382+0.017a$^\dag$ & 136.49 & 17.95 & 22.94 & 0.66 & 25.37 & 0.40 & 2.06 & C \\ 
G000.382+0.017b & 102.67 & 10.40 & 31.67 & 1.73 & 41.03 & 1.98 & 2.06 & C \\ 
G000.729$-$0.103a & 142.71 & 2.40 & 22.47 & 0.20 & 104.45 & 0.15 & 2.28 & C \\ 
G000.729$-$0.103b & 107.02 & 1.15 & 30.20 & 0.47 & 82.44 & 0.34 & 2.28 & C \\ 
G000.838+0.189 & 125.72 & 0.85 & 25.51 & 0.20 & 5.24 & 0.08 & 2.52 & A \\ 
G001.324+0.104 & 104.77 & 0.60 & 27.00 & 0.18 & $-$13.03 & 0.08 & 2.34 & A \\ 
G001.330+0.088 & 99.01 & 0.61 & 27.24 & 0.19 & $-$13.41 & 0.08 & 2.66 & A \\ 
G002.404+0.068 & 55.97 & 0.45 & 23.50 & 0.22 & 7.45 & 0.09 & 2.30 & A \\ 
G004.346+0.115 & 119.48 & 0.58 & 17.86 & 0.10 & 6.77 & 0.04 & 2.53 & A \\ 
G004.527$-$0.136 & 61.57 & 0.56 & 23.20 & 0.24 & 10.17 & 0.10 & 2.37 & A \\ 
G007.254$-$0.073a & 84.23 & 0.48 & 18.03 & 0.13 & 47.07 & 0.05 & 2.24 & B \\ 
G010.232$-$0.301 & 279.80 & 0.96 & 18.69 & 0.07 & 11.70 & 0.03 & 2.13 & A \\ 
G010.473+0.028 & 87.05 & 0.53 & 21.65 & 0.15 & 68.26 & 0.07 & 2.18 & A \\ 
G014.478$-$0.006 & 81.05 & 0.43 & 26.61 & 0.16 & 24.48 & 0.07 & 2.60 & A \\ 
G014.489+0.020 & 88.96 & 0.90 & 28.39 & 0.33 & 28.42 & 0.14 & 3.23 & A \\ 
G015.125$-$0.529 & 243.53 & 1.16 & 25.66 & 0.14 & 18.74 & 0.06 & 2.73 & A \\ 
G016.361$-$0.209 & 53.83 & 0.46 & 18.44 & 0.18 & 46.56 & 0.08 & 2.44 & A \\ 
G016.943$-$0.074 & 106.25 & 0.59 & 30.66 & 0.20 & $-$4.72 & 0.08 & 2.40 & A \\ 
G018.077+0.071 & 55.39 & 0.54 & 18.86 & 0.21 & 58.30 & 0.09 & 2.21 & A \\ 
G018.677$-$0.236a & 65.26 & 0.60 & 29.66 & 0.14 & 45.24 & 0.14 & 2.31 & B \\ 
G018.832$-$0.300 & 45.21 & 0.33 & 29.28 & 0.25 & 46.31 & 0.11 & 1.78 & A \\ 
G018.937$-$0.434a & 221.85 & 0.91 & 23.83 & 0.05 & 67.94 & 0.05 & 2.20 & A \\ 
G020.450+0.025 & 71.13 & 1.07 & 12.74 & 0.27 & 73.93 & 0.27 & 2.19 & A \\ 
G020.533$-$0.187 & 40.17 & 0.56 & 16.38 & 0.27 & 48.57 & 0.11 & 1.83 & A \\ 
G022.755$-$0.246a & 73.85 & 0.54 & 19.12 & 0.17 & 106.81 & 0.07 & 2.04 & B \\ 
G022.755$-$0.246b & 54.84 & 0.50 & 22.35 & 0.25 & 69.89 & 0.10 & 2.04 & B \\ 
G023.029$-$0.405 & 137.32 & 1.05 & 14.71 & 0.15 & 76.60 & 0.15 & 2.26 & A \\ 
G023.265$-$0.301a & 126.85 & 0.62 & 17.45 & 0.14 & 73.36 & 0.06 & 2.09 & B \\ 
G023.585+0.029 & 103.41 & 0.55 & 24.23 & 0.15 & 92.39 & 0.06 & 1.84 & A \\ 
G024.500+0.487a & 148.70 & 0.67 & 16.73 & 0.09 & 99.11 & 0.04 & 2.40 & A \\ 
G024.735+0.159a & 83.21 & 0.50 & 18.87 & 0.17 & 109.54 & 0.07 & 2.14 & B \\ 
G024.739+0.083 & 187.01 & 0.88 & 17.30 & 0.09 & 111.41 & 0.04 & 2.11 & A \\ 
G025.150+0.092a & 57.60 & 0.52 & 20.97 & 0.22 & 46.51 & 0.09 & 1.80 & A \\ 
G025.220+0.289 & 35.82 & 0.29 & 23.14 & 0.22 & 42.41 & 0.09 & 1.53 & A \\ 
G027.334+0.176a & 42.71 & 0.59 & 16.44 & 0.32 & 79.99 & 0.18 & 1.79 & B \\ 
G028.304$-$0.390 & 170.36 & 0.48 & 21.59 & 0.07 & 77.33 & 0.03 & 2.47 & A \\ 
G028.696+0.048a & 189.68 & 1.04 & 19.85 & 0.07 & 100.84 & 0.07 & 2.33 & B \\ 
G028.696+0.048b & 64.54 & 1.01 & 11.87 & 0.09 & 116.88 & 0.09 & 2.33 & C \\ 
G030.234$-$0.139a & 86.59 & 0.90 & 18.48 & 0.09 & 98.73 & 0.09 & 1.93 & B \\ 
G030.374+0.026a & 65.60 & 0.27 & 20.81 & 0.10 & 44.56 & 0.04 & 1.51 & A \\ 
G030.644+0.053a & 53.66 & 0.41 & 27.70 & 0.24 & 99.68 & 0.10 & 2.23 & A \\ 
G030.797+0.165a & 69.90 & 0.36 & 21.25 & 0.13 & 40.33 & 0.05 & 2.08 & A \\ 
G030.838+0.114a & 71.61 & 0.43 & 25.29 & 0.18 & 35.70 & 0.07 & 1.86 & A \\ 
G030.852+0.149a & 184.25 & 0.48 & 22.34 & 0.07 & 39.80 & 0.03 & 2.24 & A \\ 
G030.883+0.071a & 96.00 & 0.36 & 22.69 & 0.10 & 96.42 & 0.04 & 1.59 & A \\ 
G030.956+0.599a & 71.12 & 0.22 & 22.46 & 0.08 & 23.32 & 0.03 & 1.10 & A \\ 
G031.157$-$0.148a & 78.23 & 0.54 & 20.41 & 0.17 & 43.53 & 0.07 & 1.99 & B \\ 
G031.470$-$0.344 & 83.97 & 0.35 & 22.14 & 0.11 & 88.97 & 0.04 & 1.47 & A \\ 
G032.058+0.077 & 61.66 & 0.65 & 16.85 & 0.21 & 96.37 & 0.09 & 2.12 & A \\ 
G032.272$-$0.226 & 65.07 & 0.29 & 25.88 & 0.13 & 21.45 & 0.06 & 1.60 & A \\ 
G032.928+0.607 & 50.97 & 0.18 & 27.68 & 0.12 & $-$38.20 & 0.05 & 0.96 & A \\ 
G034.031$-$0.059a & 93.93 & 0.46 & 12.18 & 0.04 & 56.54 & 0.04 & 1.37 & A \\ 
G034.133+0.471 & 117.06 & 0.39 & 25.73 & 0.10 & 34.53 & 0.04 & 2.15 & A \\ 
G034.172+0.175 & 49.77 & 0.42 & 16.70 & 0.16 & 57.33 & 0.07 & 1.84 & A \\ 
G034.686+0.068 & 43.67 & 0.27 & 26.65 & 0.19 & 52.09 & 0.08 & 1.34 & A \\ 
G035.541+0.005 & 83.83 & 0.28 & 21.09 & 0.08 & 57.58 & 0.03 & 1.25 & A \\ 
G037.816$-$0.379 & 57.30 & 0.28 & 21.18 & 0.12 & 64.13 & 0.05 & 1.77 & A \\ 
G037.868$-$0.601 & 49.60 & 0.27 & 30.85 & 0.20 & 47.73 & 0.08 & 1.67 & A \\ 
G038.875+0.308 & 55.67 & 0.26 & 30.73 & 0.16 & $-$15.19 & 0.07 & 1.47 & A \\ 
G039.728$-$0.396 & 51.37 & 0.30 & 25.08 & 0.17 & 58.26 & 0.07 & 2.07 & A \\ 
G039.873$-$0.177 & 42.40 & 0.29 & 23.47 & 0.19 & 59.69 & 0.08 & 1.77 & A \\ 
G039.883$-$0.346 & 62.67 & 0.26 & 32.61 & 0.16 & 58.96 & 0.07 & 1.86 & A \\ 
G041.515$-$0.139 & 92.61 & 0.26 & 29.54 & 0.10 & 57.82 & 0.04 & 1.48 & A \\ 
G042.209$-$0.587 & 44.37 & 0.29 & 21.61 & 0.16 & 72.01 & 0.07 & 1.67 & A \\ 
G043.818+0.393 & 30.04 & 0.16 & 28.19 & 0.18 & $-$10.59 & 0.07 & 0.94 & A \\ 
G044.501+0.335 & 48.38 & 0.26 & 22.27 & 0.14 & $-$43.22 & 0.06 & 2.11 & A \\ 
G048.551$-$0.001 & 101.38 & 0.31 & 24.12 & 0.08 & 19.27 & 0.04 & 1.29 & A \\ 
G049.163$-$0.066 & 36.11 & 0.28 & 19.97 & 0.18 & 60.23 & 0.07 & 1.58 & A \\ 
G049.399$-$0.489 & 134.94 & 0.47 & 21.71 & 0.09 & 59.13 & 0.04 & 1.85 & A \\ 
G049.828+0.366 & 46.05 & 0.38 & 21.53 & 0.21 & 4.32 & 0.09 & 1.75 & A \\ 
G052.098+1.042 & 75.71 & 0.31 & 28.97 & 0.14 & 36.19 & 0.06 & 1.97 & A \\ 
G346.056$-$0.021a & 87.78 & 0.49 & 21.20 & 0.14 & $-$76.76 & 0.06 & 2.55 & A \\ 
G347.870+0.015 & 103.81 & 0.71 & 23.54 & 0.19 & $-$32.21 & 0.08 & 3.35 & A \\ 
G348.533$-$0.972 & 133.62 & 0.70 & 23.82 & 0.14 & $-$12.06 & 0.06 & 2.43 & A \\ 
G348.557$-$0.985 & 108.50 & 0.61 & 20.73 & 0.14 & $-$10.57 & 0.06 & 2.54 & A \\ 
G350.004+0.438 & 65.82 & 0.52 & 30.41 & 0.15 & $-$34.33 & 0.15 & 2.20 & A \\ 
G350.177+0.017 & 119.82 & 0.52 & 22.81 & 0.11 & $-$69.00 & 0.05 & 2.47 & A \\ 
G350.330+0.157 & 93.67 & 0.55 & 16.76 & 0.11 & $-$62.66 & 0.05 & 2.39 & A \\ 
G352.313$-$0.440 & 69.08 & 0.61 & 21.97 & 0.22 & $-$13.75 & 0.09 & 1.96 & A \\ 
G354.610+0.484 & 149.98 & 0.50 & 26.45 & 0.10 & $-$23.55 & 0.04 & 2.29 & A \\ 
G354.717+0.293 & 120.91 & 0.75 & 23.46 & 0.17 & 95.44 & 0.07 & 2.30 & A \\ 
G355.344+0.145 & 93.72 & 0.43 & 33.09 & 0.18 & 16.38 & 0.07 & 3.03 & A \\ 
G358.633+0.062 & 83.29 & 0.50 & 26.35 & 0.18 & 13.87 & 0.08 & 2.43 & A \\ 
G359.929+0.045a & 144.20 & 0.68 & 30.73 & 0.17 & $-$50.27 & 0.07 & 2.67 & A \\ 
G001.330+0.150 & 30.80 & 0.36 & 22.22 & 0.31 & $-$19.98 & 0.13 & 2.43 & A \\ 
G008.373$-$0.352 & 74.20 & 0.45 & 23.07 & 0.16 & 34.88 & 0.07 & 2.25 & A \\ 
G008.432$-$0.276 & 49.79 & 0.46 & 18.72 & 0.20 & 34.83 & 0.08 & 2.61 & A \\ 
G009.875$-$0.749 & 79.13 & 0.40 & 24.24 & 0.14 & 30.90 & 0.06 & 1.94 & A \\ 
G010.638$-$0.434 & 72.21 & 0.54 & 18.23 & 0.16 & $-$0.38 & 0.07 & 1.97 & A \\ 
G022.739$-$0.303a & 34.23 & 0.30 & 23.48 & 0.24 & 69.02 & 0.10 & 1.43 & B \\ 
G028.696+0.048c & 8.76 & 0.49 & 17.03 & 0.47 & 38.92 & 0.47 & 2.33 & B \\ 
G030.022$-$0.044 & 105.13 & 0.49 & 21.73 & 0.12 & 94.91 & 0.05 & 1.35 & A \\ 
G030.249+0.243a & 34.09 & 0.30 & 20.99 & 0.21 & 8.70 & 0.09 & 1.60 & A \\ 
G034.041+0.053 & 38.04 & 0.34 & 22.49 & 0.23 & 36.41 & 0.10 & 1.53 & A \\ 
G038.930$-$0.386 & 11.61 & 0.28 & 20.89 & 0.58 & 41.84 & 0.24 & 1.66 & B \\ 
G045.542$-$0.006 & 41.90 & 0.27 & 25.89 & 0.19 & 54.94 & 0.08 & 1.67 & A \\ 
G056.252$-$0.160 & 7.86 & 0.26 & 17.34 & 0.67 & 35.37 & 0.28 & 1.48 & B \\ 
G343.856$-$0.106 & 37.56 & 0.64 & 20.62 & 0.42 & $-$17.77 & 0.17 & 2.94 & B \\ 
G344.991$-$0.266 & 11.14 & 0.44 & 15.14 & 0.70 & $-$29.40 & 0.29 & 1.56 & B \\ 
G351.265+1.019 & 51.43 & 0.36 & 26.95 & 0.22 & $-$3.72 & 0.09 & 1.86 & A \\ 
G351.691+0.669 & 72.62 & 0.40 & 20.13 & 0.13 & 2.99 & 0.05 & 2.45 & A \\ 
G356.470$-$0.001 & 20.14 & 0.42 & 23.61 & 0.59 & 4.46 & 0.24 & 2.63 & B \\ 
G358.616$-$0.076 & 62.81 & 0.47 & 21.66 & 0.19 & $-$212.91 & 0.08 & 1.73 & A \\ 
G358.720+0.011 & 38.68 & 0.37 & 23.74 & 0.27 & $-$206.54 & 0.11 & 1.80 & A \\ 

\enddata

\tablenotetext{\dag\,}{Source names with an appended letter flag a
  specific emission component in an \hii region spectrum that has
  hydrogen RRLs at several different LSR velocities.  See the Paper II
  catalog.}

\label{tab:hdata}
\end{deluxetable}

\clearpage
\centering
\begin{deluxetable}{lrcrcrccc}
\tablecaption{Helium Radio Recombination Line Parameters}
\tablewidth{0pt}
\tablehead{
  \colhead{Source} &
  \colhead{$T_{\rm L}$} &
  \colhead{$\sigma\, T_{\rm L}$} &
  \colhead{$\Delta v$} &
  \colhead{$\sigma\, \Delta v$} &
  \colhead{$V_{\rm lsr}$} &
  \colhead{$\sigma\, V_{\rm lsr}$} &
  \colhead{rms} &
  \colhead{QF} \\

  \colhead{ } &
  \colhead{\mK} &
  \colhead{\mK} &
  \colhead{\kms} &
  \colhead{\kms} &
  \colhead{\kms} &
  \colhead{\kms} &
  \colhead{\mK} &
  \colhead{ }
}

\startdata
G000.382+0.017a & 11.04 & 0.42 & 28.26 & 1.73 & 25.19 & 0.56 & 2.06 & C \\ 
G000.382+0.017b & 4.89 & 0.63 & 11.54 & 1.88 & 53.54 & 0.78 & 2.06 & C \\ 
G000.729$-$0.103a & 13.33 & 0.81 & 15.01 & 1.19 & 103.15 & 0.57 & 2.28 & C \\ 
G000.729$-$0.103b & 11.47 & 0.52 & 25.14 & 2.51 & 79.65 & 0.92 & 2.28 & C \\ 
G000.838+0.189 & 7.22 & 0.48 & 15.59 & 1.20 & 8.42 & 0.51 & 2.52 & B \\ 
G001.324+0.104 & 7.55 & 0.43 & 26.94 & 1.76 & $-$16.19 & 0.75 & 2.34 & A \\ 
G001.330+0.088 & 8.07 & 0.54 & 23.30 & 1.80 & $-$14.33 & 0.76 & 2.66 & B \\ 
G002.404+0.068 & 4.70 & 0.40 & 23.94 & 2.73 & 9.14 & 1.05 & 2.30 & C \\ 
G004.346+0.115 & 11.16 & 0.54 & 15.64 & 0.88 & 7.34 & 0.37 & 2.53 & B \\ 
G004.527$-$0.136 & 5.13 & 0.39 & 30.88 & 2.72 & 10.37 & 1.15 & 2.37 & C \\ 
G007.254$-$0.073a & 9.03 & 0.84 & 4.73 & 0.51 & 46.50 & 0.21 & 2.24 & C \\ 
G010.232$-$0.301 & 15.46 & 0.68 & 10.98 & 0.56 & 11.12 & 0.24 & 2.13 & A \\ 
G010.473+0.028 & 7.65 & 0.34 & 30.77 & 2.96 & 71.78 & 1.34 & 2.18 & B \\ 
G014.478$-$0.006 & 6.25 & 0.41 & 24.26 & 1.86 & 23.90 & 0.79 & 2.60 & B \\ 
G014.489+0.020 & 10.50 & 0.56 & 26.69 & 2.45 & 29.33 & 0.96 & 3.23 & B \\ 
G015.125$-$0.529 & 29.83 & 0.74 & 18.37 & 0.53 & 17.42 & 0.22 & 2.73 & A \\ 
G016.361$-$0.209 & 5.03 & 0.87 & 9.30 & 1.85 & 43.74 & 0.78 & 2.44 & B \\ 
G016.943$-$0.074 & 10.72 & 0.40 & 29.80 & 1.55 & $-$5.67 & 0.59 & 2.40 & A \\ 
G018.077+0.071 & 6.48 & 0.61 & 12.20 & 1.32 & 56.99 & 0.56 & 2.21 & B \\ 
G018.677$-$0.236a & 5.08 & 0.26 & 31.47 & 1.88 & 40.96 & 0.79 & 2.31 & C \\ 
G018.832$-$0.300 & 4.90 & 0.28 & 34.68 & 4.09 & 47.24 & 1.77 & 1.78 & C \\ 
G018.937$-$0.434a & 8.35 & 0.38 & 16.31 & 0.85 & 67.75 & 0.36 & 2.20 & A \\ 
G020.450+0.025 & 5.17 & 0.62 & 10.90 & 1.50 & 69.64 & 0.64 & 2.19 & C \\ 
G020.533$-$0.187 & 4.66 & 0.58 & 11.46 & 1.64 & 49.50 & 0.69 & 1.83 & C \\ 
G022.755$-$0.246a & 4.81 & 0.33 & 23.85 & 1.93 & 110.36 & 0.81 & 2.04 & C \\ 
G022.755$-$0.246b & 4.19 & 0.53 & 9.59 & 1.41 & 76.05 & 0.59 & 2.04 & C \\ 
G023.029$-$0.405 & 10.32 & 0.53 & 12.38 & 0.73 & 75.67 & 0.31 & 2.26 & B \\ 
G023.265$-$0.301a & 6.24 & 0.61 & 8.48 & 0.96 & 75.03 & 0.41 & 2.09 & C \\ 
G023.585+0.029 & 6.92 & 0.46 & 19.74 & 1.53 & 89.21 & 0.65 & 1.84 & B \\ 
G024.500+0.487a & 9.55 & 0.55 & 16.83 & 1.12 & 98.55 & 0.48 & 2.40 & B \\ 
G024.735+0.159a & 5.78 & 0.55 & 13.13 & 1.44 & 110.09 & 0.61 & 2.14 & B \\ 
G024.739+0.083 & 12.72 & 0.72 & 9.03 & 0.59 & 110.47 & 0.25 & 2.11 & A \\ 
G025.150+0.092a & 5.16 & 0.44 & 21.92 & 2.15 & 48.93 & 0.91 & 1.80 & C \\ 
G025.220+0.289 & 4.07 & 0.30 & 14.95 & 1.28 & 43.71 & 0.54 & 1.53 & B \\ 
G027.334+0.176a & 5.64 & 0.25 & 18.36 & 0.93 & 80.54 & 0.40 & 1.79 & C \\ 
G028.304$-$0.390 & 12.64 & 0.66 & 12.67 & 0.77 & 76.01 & 0.33 & 2.47 & A \\ 
G028.696+0.048a & 20.06 & 0.83 & 16.87 & 1.35 & 100.27 & 0.84 & 2.33 & C \\ 
G028.696+0.048b & 8.74 & 1.46 & 13.53 & 2.25 & 114.81 & 1.47 & 2.33 & C \\ 
G030.234$-$0.139a & 11.53 & 0.49 & 7.93 & 0.39 & 94.81 & 0.17 & 1.93 & C \\ 
G030.374+0.026a & 7.80 & 0.49 & 7.34 & 0.53 & 43.61 & 0.23 & 1.51 & A \\ 
G030.644+0.053a & 5.35 & 0.37 & 20.67 & 1.88 & 105.82 & 0.73 & 2.23 & C \\ 
G030.797+0.165a & 7.41 & 0.53 & 15.72 & 1.30 & 41.26 & 0.55 & 2.08 & B \\ 
G030.838+0.114a & 8.95 & 0.47 & 21.54 & 1.72 & 34.76 & 0.65 & 1.86 & B \\ 
G030.852+0.149a & 11.18 & 0.44 & 19.09 & 0.88 & 39.99 & 0.37 & 2.24 & A \\ 
G030.883+0.071a & 4.70 & 0.36 & 21.79 & 1.95 & 96.22 & 0.83 & 1.59 & A \\ 
G030.956+0.599a & 6.39 & 0.30 & 14.56 & 0.99 & 24.24 & 0.38 & 1.10 & A \\ 
G031.157$-$0.148a & 7.06 & 0.33 & 20.32 & 1.13 & 42.22 & 0.46 & 1.99 & B \\ 
G031.470$-$0.344 & 7.17 & 0.32 & 17.77 & 0.92 & 89.52 & 0.39 & 1.47 & A \\ 
G032.058+0.077 & 7.33 & 0.61 & 6.39 & 0.62 & 96.54 & 0.26 & 2.12 & C \\ 
G032.272$-$0.226 & 7.99 & 0.35 & 14.93 & 0.76 & 21.99 & 0.32 & 1.60 & A \\ 
G032.928+0.607 & 6.33 & 0.20 & 19.49 & 0.96 & $-$38.40 & 0.37 & 0.96 & A \\ 
G034.031$-$0.059a & 11.25 & 0.43 & 11.55 & 0.51 & 56.72 & 0.22 & 1.37 & A \\ 
G034.133+0.471 & 10.81 & 0.55 & 17.50 & 1.03 & 36.20 & 0.44 & 2.15 & A \\ 
G034.172+0.175 & 4.78 & 0.61 & 13.97 & 2.07 & 59.36 & 0.88 & 1.84 & C \\ 
G034.686+0.068 & 3.10 & 0.23 & 25.61 & 3.52 & 47.04 & 1.42 & 1.34 & C \\ 
G035.541+0.005 & 7.06 & 0.29 & 14.72 & 0.69 & 58.51 & 0.29 & 1.25 & A \\ 
G037.816$-$0.379 & 4.98 & 0.47 & 10.44 & 1.47 & 60.21 & 0.50 & 1.77 & B \\ 
G037.868$-$0.601 & 3.58 & 0.25 & 37.95 & 3.20 & 43.07 & 1.30 & 1.67 & C \\ 
G038.875+0.308 & 4.71 & 0.30 & 19.86 & 1.46 & $-$16.50 & 0.62 & 1.47 & A \\ 
G039.728$-$0.396 & 7.22 & 0.62 & 8.63 & 0.85 & 55.08 & 0.36 & 2.07 & B \\ 
G039.873$-$0.177 & 3.65 & 0.37 & 28.47 & 6.97 & 56.51 & 4.08 & 1.77 & C \\ 
G039.883$-$0.346 & 3.79 & 0.25 & 30.30 & 2.27 & 56.16 & 0.96 & 1.86 & C \\ 
G041.515$-$0.139 & 5.00 & 0.23 & 37.19 & 2.03 & 51.28 & 0.83 & 1.48 & B \\ 
G042.209$-$0.587 & 6.33 & 0.44 & 12.08 & 0.96 & 75.17 & 0.41 & 1.67 & B \\ 
G043.818+0.393 & 3.35 & 0.18 & 29.11 & 1.77 & $-$12.65 & 0.75 & 0.94 & C \\ 
G044.501+0.335 & 7.90 & 0.57 & 8.87 & 0.74 & $-$44.16 & 0.31 & 2.11 & B \\ 
G048.551$-$0.001 & 9.40 & 0.28 & 17.56 & 0.60 & 19.82 & 0.26 & 1.29 & A \\ 
G049.163$-$0.066 & 3.46 & 0.32 & 17.30 & 1.83 & 65.99 & 0.78 & 1.58 & C \\ 
G049.399$-$0.489 & 8.67 & 0.27 & 33.08 & 1.18 & 63.79 & 0.50 & 1.85 & A \\ 
G049.828+0.366 & 3.64 & 0.32 & 21.03 & 3.21 & 6.36 & 1.28 & 1.75 & C \\ 
G052.098+1.042 & 7.33 & 0.39 & 22.94 & 3.00 & 30.40 & 1.53 & 1.97 & B \\ 
G346.056$-$0.021a & 6.18 & 0.40 & 32.27 & 2.42 & $-$81.79 & 1.02 & 2.55 & C \\ 
G347.870+0.015 & 8.97 & 0.70 & 17.81 & 1.60 & $-$33.15 & 0.68 & 3.35 & B \\ 
G348.533$-$0.972 & 9.79 & 0.36 & 33.83 & 2.23 & $-$13.74 & 0.90 & 2.43 & C \\ 
G348.557$-$0.985 & 12.03 & 0.25 & 39.28 & 1.49 & $-$7.77 & 0.59 & 2.54 & C \\ 
G350.004+0.438 & 5.53 & 0.57 & 9.01 & 1.07 & $-$35.25 & 0.45 & 2.20 & B \\ 
G350.177+0.017 & 9.10 & 0.64 & 14.97 & 1.21 & $-$66.57 & 0.52 & 2.47 & C \\ 
G350.330+0.157 & 8.78 & 0.54 & 12.87 & 0.92 & $-$63.96 & 0.39 & 2.39 & B \\ 
G352.313$-$0.440 & 6.38 & 0.41 & 15.10 & 1.17 & $-$9.83 & 0.48 & 1.96 & B \\ 
G354.610+0.484 & 10.79 & 0.53 & 25.39 & 2.08 & $-$25.61 & 0.81 & 2.29 & A \\ 
G354.717+0.293 & 15.20 & 0.57 & 18.85 & 0.82 & 94.13 & 0.35 & 2.30 & A \\ 
G355.344+0.145 & 7.15 & 0.47 & 36.85 & 5.98 & 21.78 & 2.98 & 3.03 & B \\ 
G358.633+0.062 & 5.76 & 0.43 & 15.87 & 1.36 & 8.45 & 0.58 & 2.43 & C \\ 
G359.929+0.045a & 13.60 & 0.68 & 16.73 & 0.97 & $-$50.46 & 0.41 & 2.67 & A \\ 

\enddata

\label{tab:hedata}
\end{deluxetable}

\clearpage
\centering
\begin{deluxetable}{lrcrcrccc}
\tablecaption{Carbon Radio Recombination Line Parameters}
\tablewidth{0pt}
\tablehead{
  \colhead{Source} &
  \colhead{$T_{\rm L}$} &
  \colhead{$\sigma\, T_{\rm L}$} &
  \colhead{$\Delta v$} &
  \colhead{$\sigma\, \Delta v$} &
  \colhead{$V_{\rm lsr}$} &
  \colhead{$\sigma\, V_{\rm lsr}$} &
  \colhead{rms} &
  \colhead{QF} \\

  \colhead{ } &
  \colhead{\mK} &
  \colhead{\mK} &
  \colhead{\kms} &
  \colhead{\kms} &
  \colhead{\kms} &
  \colhead{\kms} &
  \colhead{\mK} &
  \colhead{ }
}

\startdata
G001.330+0.088 & 7.14 & 0.95 & 2.92 & 0.45 & $-$8.18 & 0.19 & 2.66 & B \\ 
G001.330+0.150$^\dag$ & 9.75 & 0.96 & 3.62 & 0.41 & $-$16.25 & 0.18 & 2.43 & A \\ 
G008.373$-$0.352$^\dag$ & 8.07 & 0.78 & 4.32 & 0.48 & 35.64 & 0.21 & 2.25 & A \\ 
G008.432$-$0.276$^\dag$ & 14.89 & 1.11 & 3.59 & 0.31 & 36.69 & 0.13 & 2.61 & A \\ 
G009.875$-$0.749$^\dag$ & 14.96 & 0.65 & 6.88 & 0.34 & 27.55 & 0.15 & 1.94 & A \\ 
G010.473+0.028 & 5.71 & 0.53 & 12.32 & 1.31 & 69.35 & 0.56 & 2.18 & B \\ 
G010.638$-$0.434$^\dag$ & 6.76 & 0.45 & 11.06 & 0.85 & $-$3.76 & 0.36 & 1.97 & A \\ 
G014.489+0.020 & 7.21 & 1.14 & 6.38 & 1.16 & 24.66 & 0.49 & 3.23 & B \\ 
G015.125$-$0.529 & 14.21 & 1.19 & 2.79 & 0.27 & 21.70 & 0.11 & 2.73 & A \\ 
G016.361$-$0.209 & 9.21 & 0.77 & 5.22 & 0.50 & 48.77 & 0.21 & 2.44 & A \\ 
G016.943$-$0.074 & 10.01 & 0.61 & 7.41 & 0.53 & $-$5.12 & 0.22 & 2.40 & A \\ 
G018.077+0.071 & 5.48 & 0.50 & 6.45 & 0.68 & 50.84 & 0.29 & 2.21 & B \\ 
G022.739$-$0.303a$^\dag$ & 3.61 & 0.32 & 10.41 & 1.07 & 67.14 & 0.45 & 1.43 & B \\ 
G023.585+0.029 & 6.49 & 1.16 & 3.16 & 0.65 & 86.41 & 0.28 & 1.84 & C \\ 
G024.739+0.083 & 7.84 & 0.90 & 4.20 & 0.56 & 109.53 & 0.24 & 2.11 & B \\ 
G028.304$-$0.390 & 10.82 & 0.51 & 3.99 & 0.22 & 84.40 & 0.09 & 2.47 & A \\ 
G028.696+0.048b & 9.92 & 0.39 & 2.65 & 0.13 & 98.58 & 0.05 & 2.33 & B \\ 
G028.696+0.048c$^\dag$ & 13.91 & 0.42 & 4.66 & 0.38 & 38.53 & 0.10 & 2.33 & C \\ 
G030.022$-$0.044$^\dag$ & 7.67 & 0.52 & 3.99 & 0.32 & 93.40 & 0.13 & 1.35 & A \\ 
G030.249+0.243a$^\dag$ & 4.07 & 0.35 & 6.61 & 0.65 & 0.85 & 0.28 & 1.60 & B \\ 
G030.838+0.114a & 5.04 & 0.52 & 10.00 & 1.25 & 37.56 & 0.51 & 1.86 & B \\ 
G030.883+0.071a & 5.46 & 1.03 & 2.06 & 0.45 & 98.26 & 0.19 & 1.59 & C \\ 
G030.956+0.599a & 3.48 & 0.22 & 14.90 & 1.42 & 16.59 & 0.48 & 1.10 & B \\ 
G031.157$-$0.148a & 7.89 & 0.56 & 8.91 & 0.73 & 39.80 & 0.31 & 1.99 & A \\ 
G032.058+0.077 & 4.89 & 1.02 & 2.41 & 0.58 & 103.81 & 0.25 & 2.12 & C \\ 
G032.928+0.607 & 4.29 & 0.26 & 11.45 & 0.81 & $-$35.98 & 0.34 & 0.96 & B \\ 
G034.031$-$0.059a & 2.79 & 0.33 & 12.80 & 1.75 & 60.26 & 0.75 & 1.37 & C \\ 
G034.041+0.053$^\dag$ & 3.89 & 0.54 & 5.95 & 0.95 & 37.88 & 0.40 & 1.53 & B \\ 
G034.133+0.471 & 7.62 & 0.74 & 3.85 & 0.56 & 34.51 & 0.21 & 2.15 & B \\ 
G034.686+0.068 & 5.67 & 0.41 & 9.05 & 0.75 & 47.22 & 0.32 & 1.34 & B \\ 
G038.875+0.308 & 2.96 & 0.28 & 15.06 & 1.64 & $-$20.79 & 0.69 & 1.47 & B \\ 
G038.930$-$0.386$^\dag$ & 9.93 & 0.67 & 4.95 & 0.39 & 39.21 & 0.16 & 1.66 & A \\ 
G042.209$-$0.587 & 4.53 & 0.55 & 3.87 & 0.55 & 69.01 & 0.23 & 1.67 & A \\ 
G045.542$-$0.006$^\dag$ & 4.06 & 0.42 & 9.92 & 1.20 & 53.80 & 0.51 & 1.67 & B \\ 
G049.828+0.366 & 4.46 & 0.27 & 21.45 & 1.52 & 6.54 & 0.64 & 1.75 & B \\ 
G052.098+1.042 & 10.15 & 1.20 & 2.91 & 0.40 & 40.28 & 0.17 & 1.97 & A \\ 
G056.252$-$0.160$^\dag$ & 9.03 & 0.60 & 4.32 & 0.33 & 34.98 & 0.14 & 1.48 & A \\ 
G343.856$-$0.106$^\dag$ & 17.54 & 1.00 & 5.71 & 0.38 & $-$24.12 & 0.16 & 2.94 & A \\ 
G344.991$-$0.266$^\dag$ & 11.25 & 0.65 & 5.29 & 0.35 & $-$28.65 & 0.15 & 1.56 & A \\ 
G348.533$-$0.972 & 6.00 & 0.54 & 16.55 & 1.72 & $-$15.42 & 0.73 & 2.43 & C \\ 
G348.557$-$0.985 & 12.00 & 0.48 & 13.33 & 0.62 & $-$15.05 & 0.26 & 2.54 & B \\ 
G350.004+0.438 & 7.86 & 0.73 & 3.61 & 0.39 & $-$30.97 & 0.17 & 2.20 & B \\ 
G350.177+0.017 & 6.60 & 0.85 & 5.49 & 0.82 & $-$69.74 & 0.35 & 2.47 & C \\ 
G351.265+1.019$^\dag$ & 11.06 & 0.58 & 5.18 & 0.31 & 0.76 & 0.13 & 1.86 & A \\ 
G351.691+0.669$^\dag$ & 19.05 & 1.51 & 3.10 & 0.28 & $-$2.98 & 0.12 & 2.45 & A \\ 
G352.313$-$0.440 & 7.75 & 0.59 & 10.30 & 0.91 & $-$12.26 & 0.39 & 1.96 & B \\ 
G354.610+0.484 & 10.74 & 0.72 & 4.41 & 0.34 & $-$22.08 & 0.15 & 2.29 & A \\ 
G354.717+0.293 & 5.86 & 0.78 & 8.91 & 1.38 & 95.80 & 0.58 & 2.30 & C \\ 
G356.470$-$0.001$^\dag$ & 14.04 & 1.13 & 5.38 & 0.50 & $-$1.52 & 0.21 & 2.63 & A \\ 
G358.616$-$0.076$^\dag$ & 4.19 & 0.51 & 7.77 & 1.09 & $-$208.07 & 0.46 & 1.73 & B \\ 
G358.720+0.011$^\dag$ & 6.07 & 0.44 & 8.53 & 0.71 & $-$215.13 & 0.30 & 1.80 & A \\ 
G359.929+0.045a & 8.61 & 1.52 & 3.00 & 0.61 & $-$51.40 & 0.26 & 2.67 & C \\ 

\enddata

\tablenotetext{\dag\,}{Has carbon, but no helium, RRL emission.}

\label{tab:cdata}
\end{deluxetable}

\centering
\begin{deluxetable}{ccccc}
\rotate
\tablecaption{Summary of HRDS Radio Recombination Line Properties}
\tablewidth{0pt}
\tablehead{
  \colhead{X} &
  \colhead{$\rm <T_L(X)/T_L(H)>$} &
  \colhead{$\rm <\Delta v(X)/\Delta v(H)>$} &
  \colhead{$\rm <V_{lsr}(X)-V_{lsr}(H)>$} &
  \colhead{$\rm <\,\vert\,V_{lsr}(X)-V_{lsr}(H)\,\vert\,>$} \\
  \colhead{} &
  \colhead{\%} &
  \colhead{\%} &
  \colhead{\kms} &
  \colhead{\kms} 
}

\startdata
He & 
\phn 9.1 $\pm$ \phn 2.6 &
76.6 $\pm$ 24.7 &
$-$0.34 $\pm$ 2.29 &
1.69 $\pm$ 1.57 \\

C &
20.0 $\pm$ 25.5 &
31.4 $\pm$ 18.1 &
$-$1.30 $\pm$ 4.79 &
3.78 $\pm$ 3.18 \\

C with no He &
33.6 $\pm$ 33.9 &
28.6 $\pm$ 11.3 &
$-$1.64 $\pm$ 4.06 &
3.52 $\pm$ 2.50 \\

C with He &
\phn 9.2 $\pm$ \phn 3.3 &
33.7 $\pm$ 22.0 &
$-$1.04 $\pm$ 5.37 &
3.98 $\pm$ 3.67 \\

\enddata

\label{tab:rrlproperties}
\end{deluxetable}

\centering
\begin{deluxetable}{cccccc}
\rotate
\tablecaption{Comparison of Carbon RRL Emission Properties and MIR Morphology}
\tablewidth{0pt}
\tablehead{
  \colhead{Morphology\tablenotemark{a}} &
  \colhead{N} &
  \colhead{$\rm <T_L(C)/T_L(H)>$} &
  \colhead{$\rm <\Delta v(C)/\Delta v(H)>$} &
  \colhead{$\rm <V_{lsr}(C)-V_{lsr}(H)>$} &
  \colhead{$\rm <\,\vert\,V_{lsr}(C)-V_{lsr}(H)\,\vert\,>$} \\
  \colhead{ } &
  \colhead{ } &
  \colhead{\%} &
  \colhead{\%} &
  \colhead{\kms} &
  \colhead{\kms} 
}

\startdata
B  &     10 & 22.5 $\pm$ 25.1     & 35.9 $\pm$     27.2 &   $-$1.65 $\pm$ 4.39 & 4.11 $\pm$ 1.89 \\
BB & \phn 2 & 60.7 $\pm$ 76.6     & 20.0 $\pm$ \phn 7.1 & \phs 1.68 $\pm$ 0.48 & 1.68 $\pm$ 0.48 \\
PB & \phn 2 & 10.1 $\pm$ \phn 1.1 & 39.7 $\pm$     29.7 &   $-$1.32 $\pm$ 2.93 & 2.07 $\pm$ 1.86 \\
IB & \phn 9 & 14.0 $\pm$ \phn 9.7 & 23.8 $\pm$     17.0 &   $-$2.21 $\pm$ 7.37 & 5.43 $\pm$ 5.16 \\
C  & \phn 8 & 22.7 $\pm$ 31.9     & 35.8 $\pm$ \phn 7.6 &   $-$2.62 $\pm$ 3.62 & 3.36 $\pm$ 2.84 \\
I  &     12 & 15.3 $\pm$ 18.3     & 31.1 $\pm$     14.0 & \phs 0.36 $\pm$ 4.23 & 3.41 $\pm$ 2.31 \\

\enddata

\tablenotetext{a}{Morphological source structure classification based on {\it Spitzer} 
GLIMPSE 8\micron\ images:\\  
B -- Bubble: 8\,\micron\ emission surrounding 24\,\micron\ and radio continuum emission\\
BB -- Bipolar Bubble: two bubbles connected by a region of strong MIR and radio continuum emission\\
PB -- Partial Bubble:  similar to ``B'' but not complete\\
IB -- Irregular Bubble:  similar to ``B'' but with less well-defined structure\\ 
C -- Compact:  resolved 8\,\micron\ emission with no hole in the center\\
I -- Irregular:  complex morphology not easily classified\\ }

\label{tab:c_morph}
\end{deluxetable}








\clearpage

\clearpage

\clearpage

\clearpage
\clearpage
\begin{figure}
\centering
\includegraphics[scale=0.33,angle=90]{he_peak.epsi}
\includegraphics[scale=0.33,angle=90]{he_width.epsi}
\includegraphics[scale=0.33,angle=90]{he_vlsr.epsi}
\includegraphics[scale=0.33,angle=90]{he_abs_vlsr.epsi}
\caption{Histograms comparing the helium and hydrogen RRL
  properties. Shown are the distributions of the He/H ratios of line
  intensities (top-left) and FWHM line widths (top-right), together
  with the difference between the helium and hydrogen RRL LSR
  velocities (bottom-left), and absolute value of the difference in
  the RRL velocities (bottom-right). Only helium quality factor A and
  B emission components are included.}
\label{fig:He_params}
\end{figure}

\clearpage
\begin{figure}
\centering
\includegraphics[scale=0.33,angle=90]{c_peak.epsi}
\includegraphics[scale=0.33,angle=90]{c_width.epsi}
\includegraphics[scale=0.33,angle=90]{c_vlsr.epsi}
\includegraphics[scale=0.33,angle=90]{c_abs_vlsr.epsi}
\caption{Histograms comparing the carbon and hydrogen RRL
  properties. Shown are the distributions of the C/H ratios of line
  intensities (top-left) and FWHM line widths (top-right), together
  with the difference between the carbon and hydrogen RRL LSR
  velocities (bottom-left), and absolute value of the difference in
  the RRL velocities (bottom-right).  Only carbon quality factor A
  and B emission components are included.}
 \label{fig:C_params}
\end{figure}

\clearpage
\begin{figure}
\centering
\includegraphics[scale=0.70,angle=90]{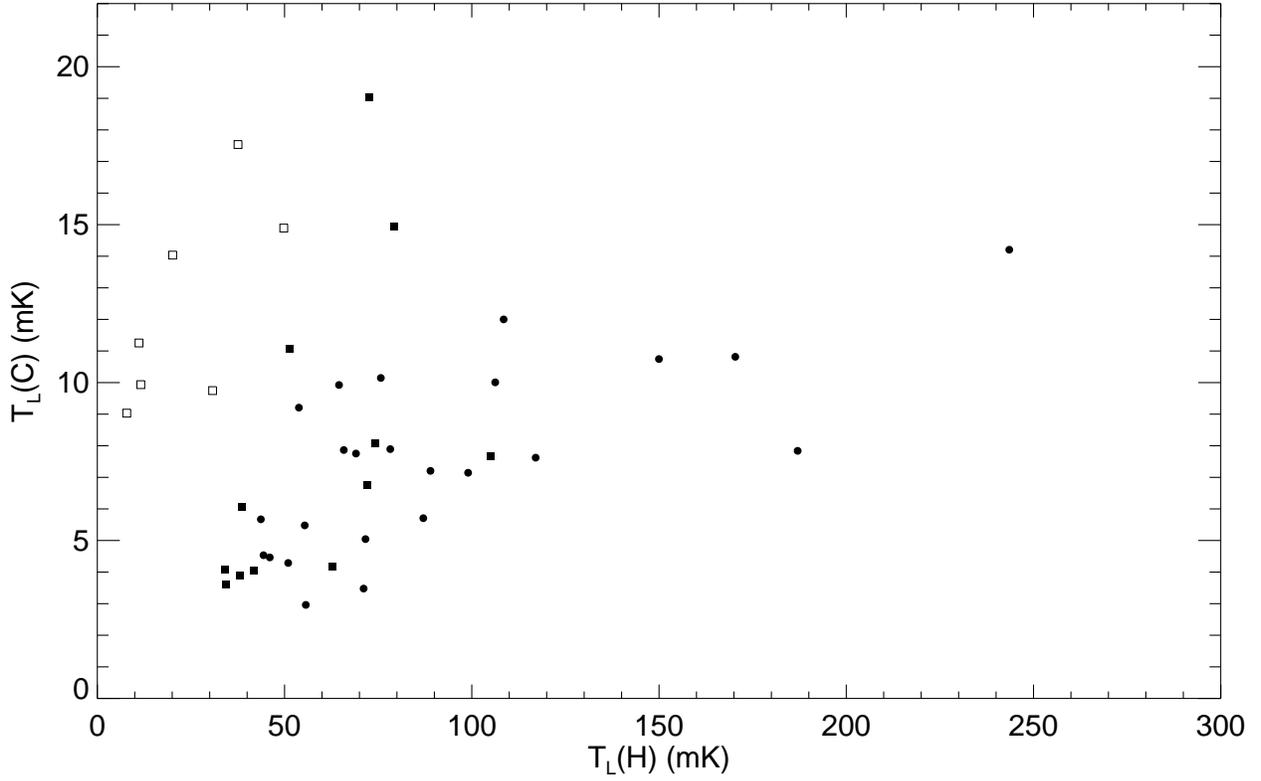}
\caption{Comparison of carbon and hydrogen RRL intensities for HRDS
  \hii regions.  The hydrogen intensity is a proxy for the nebular
  thermal continuum (see text).  Circles are nebulae that have both
  carbon and helium emission; squares are sources with only carbon
  emission. Open squares denote \hii regions that are not expected to
  have detectable helium emission based on their hydrogen RRL
  intensities.  Only carbon RRL emission components with quality
  factor A and B are shown.}
\label{fig:tc_vs_th}
\end{figure}

\clearpage
\begin{figure}
\centering
\includegraphics[scale=0.70,angle=90]{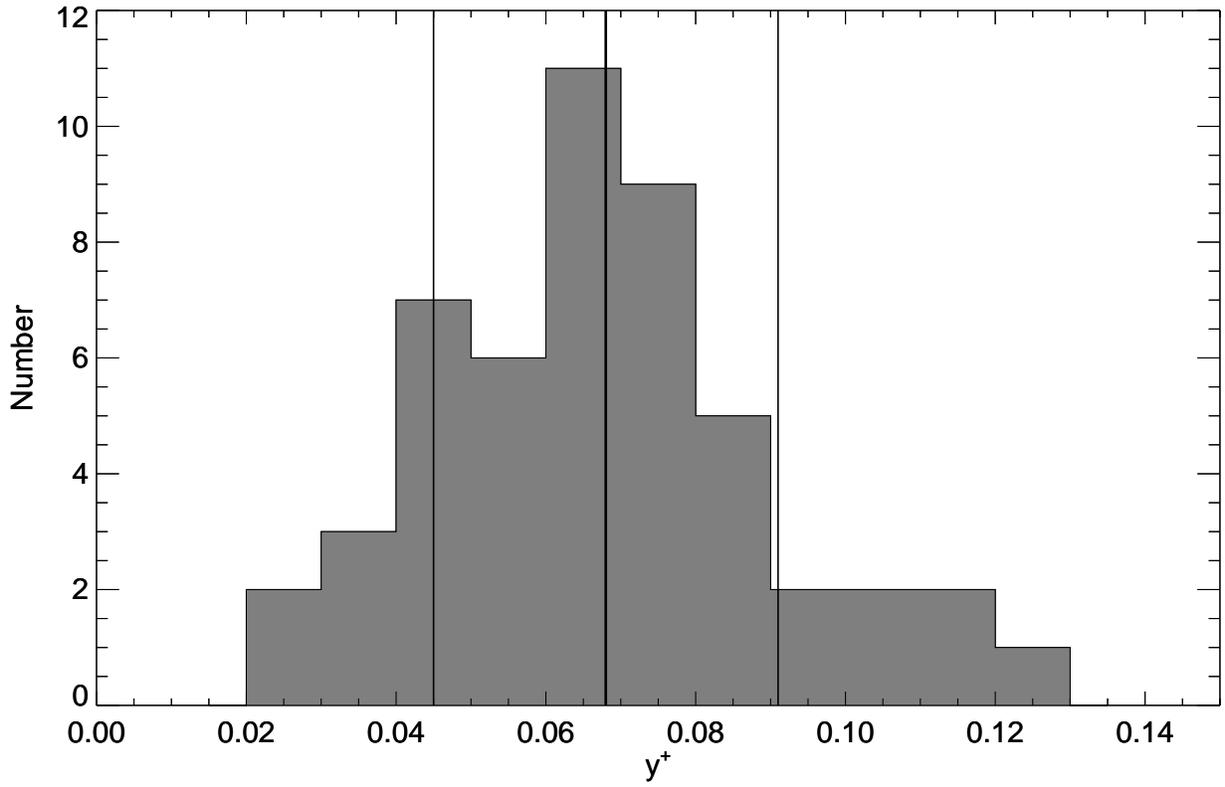}
\caption{Histogram of the $\rm \he4^+/H^+$ abundance ratio by number,
  $y^+$, for HRDS \hii regions.  Only quality factor A and B emission
  components are plotted. The vertical flags mark the mean and
  standard deviation of this distribution, $< y^+
  >\,=\,0.068\,\pm\,0.023$.}
\label{fig:yp_hist}
\end{figure}

\clearpage
\begin{figure}
\centering
\includegraphics[scale=0.70,angle=90]{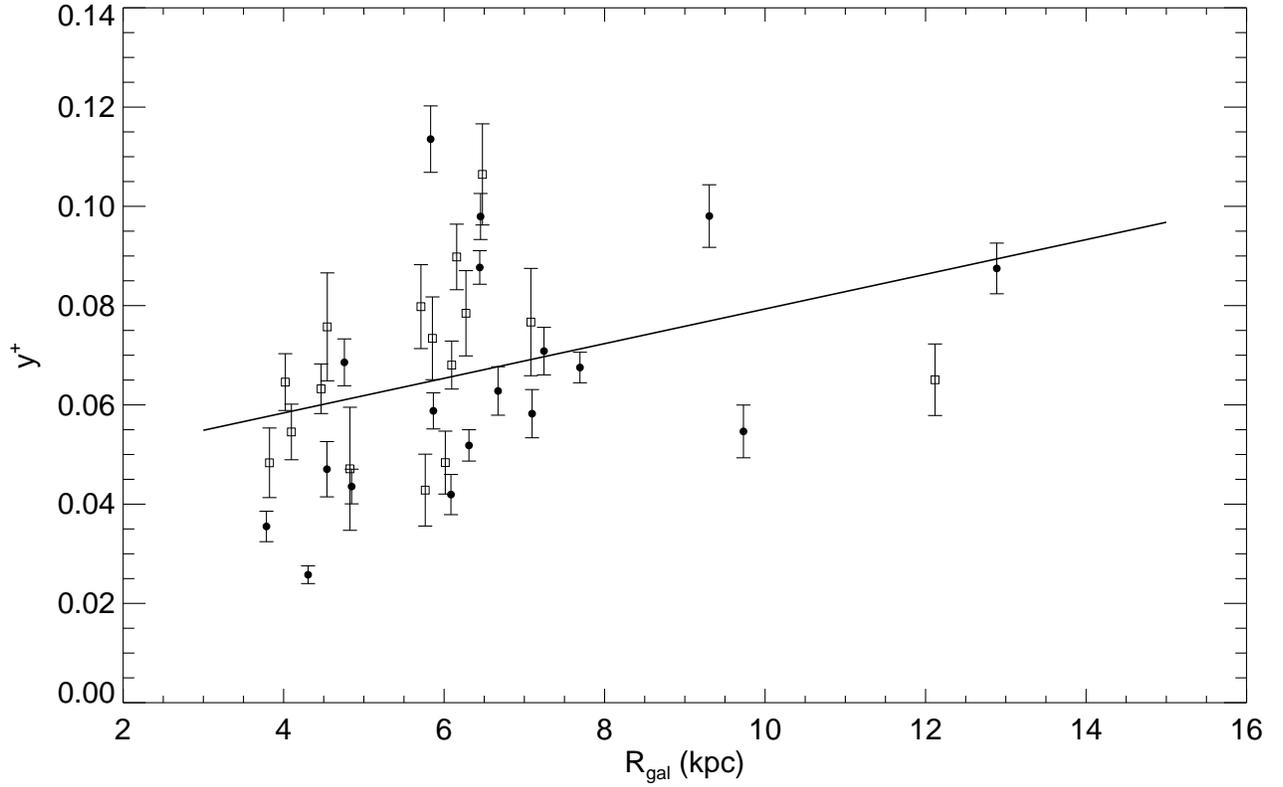}
\caption{The $\rm \he4^+/H^+$ abundance ratio by number, $y^+$, as a
  function of Galactocentric radius, $\rm R_{gal}$. Only quality
  factor A (filled circles) and B (open squares) emission components are
  shown. The solid line is an unweighted least-squares linear fit to
  the data. Error bars are $\pm\,1\,\sigma$ uncertainties in $y^+$.}
\label{fig:rgal}
\end{figure}

\clearpage
\begin{figure}
\centering
\includegraphics[scale=0.70,angle=90]{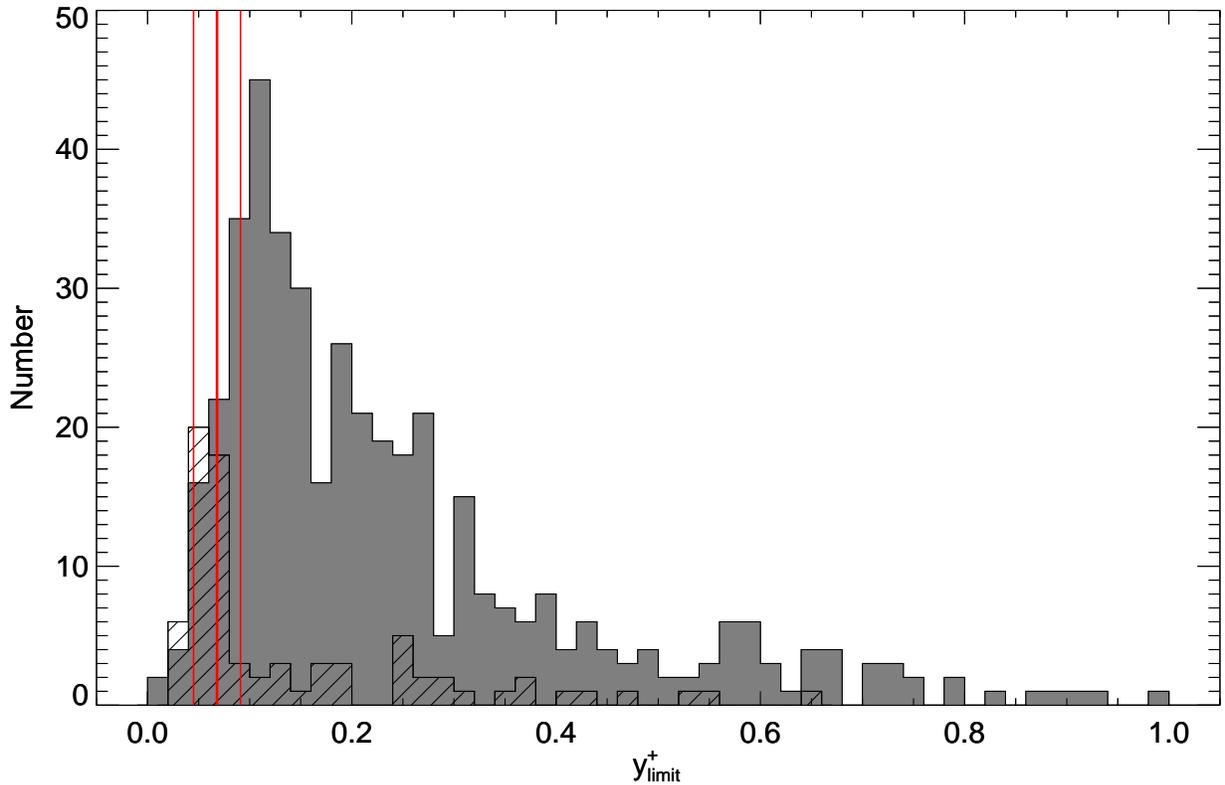}
\caption{Histograms of upper limits for the $\rm \he4^+/H^+$ abundance
  ratio by number, $y_{\rm limit}^+$.  The black hatched histogram is
  the $y_{\rm limit}^+$ distribution for the 79 nebulae with the most
  astrophysically meaningful limits and the grey histogram is the
  distribution for the remaining HRDS RRL emission components (see
  Section~\ref{sec:disc_limits}).  The red vertical flags mark the
  location of the mean and standard deviation of the average abundance
  ratio for HRDS \hii regions (see Fig.~\ref{fig:yp_hist} and
  Sec.~\ref{sec:disc_abundance}).  We find 12 nebulae with very low
  \hepr4 limits, $y_{\rm limit}^+ < 0.04$}
\label{fig:all_limits}
\end{figure}

\clearpage
\begin{figure}
\centering
\includegraphics[scale=0.70,angle=90]{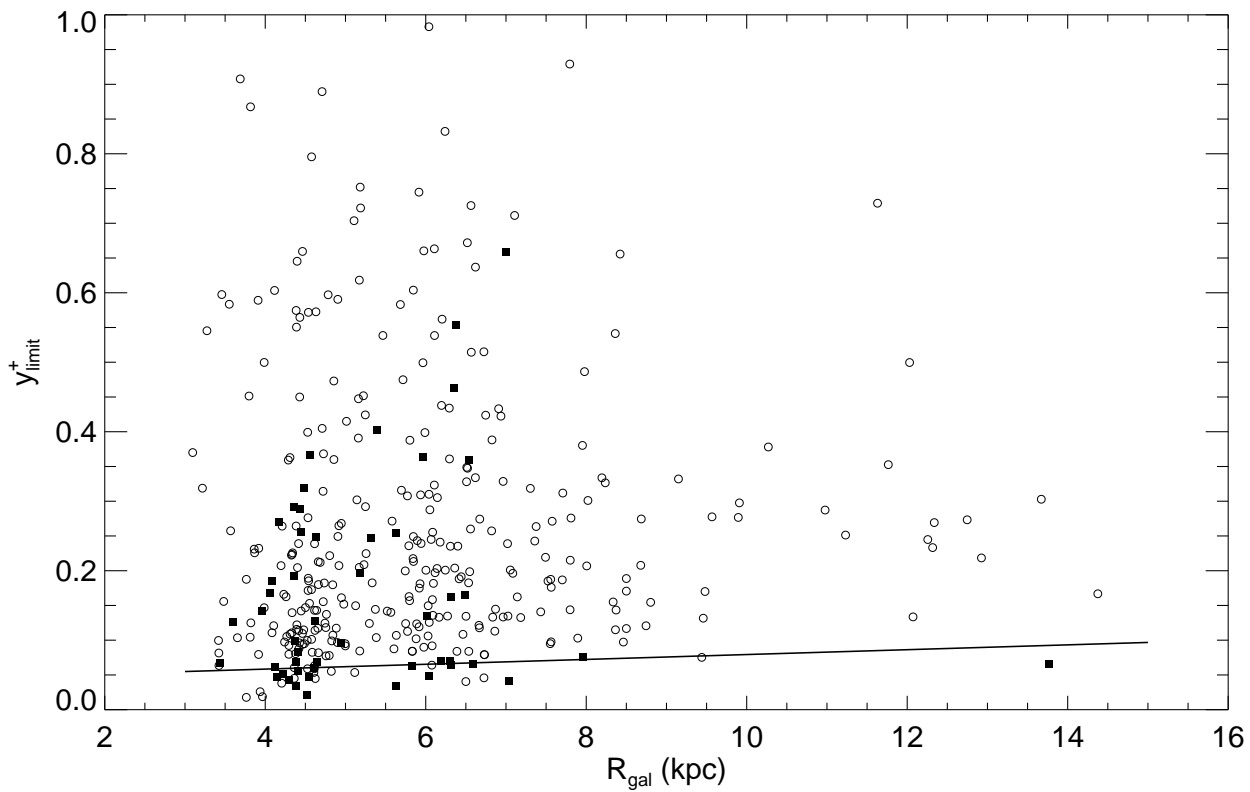}
\caption{Upper limits for the $\rm \he4^+/H^+$ abundance ratio
  by number, $y_{\rm limit}^+$, as a function of Galactocentric
  radius, \rgal.  Uncertainties are omitted for clarity.  Black
  squares show the nebulae with the most astrophysically meaningful
  limits and the open circles are the remaining limits for HRDS
  nebulae (see Sec.~\ref{sec:disc_limits}).  The solid line is the
  Figure~\ref{fig:rgal} fit to the \hii region sample with detected
  helium and hydrogen RRLs.}
\label{fig:ylimit_rgal}
\end{figure}

\end{document}